\documentclass[twocolumn,showpacs,preprintnumbers,superscriptaddress,prl,amsmath,amssymb,nofootinbib]{revtex4-1}

\usepackage{graphicx}
\usepackage{dcolumn}
\usepackage{bm}
\usepackage{color}
\usepackage{ulem}
\usepackage{gensymb}
\usepackage{braket}
\usepackage{epstopdf}
\usepackage{amsmath}
\usepackage[percent]{overpic}

\begin{document}

\title{A-type antiferromagnetic order in MnBi$_4$Te$_7$ and MnBi$_6$Te$_{10}$ single crystals\\}

\author{J.-Q. Yan}
\email{yanj@ornl.gov}
\affiliation{Materials Science and Technology Division, Oak Ridge National Laboratory, Oak Ridge, Tennessee 37831, USA}
\affiliation{Department of Materials Science and Engineering, University of Tennessee, Knoxville, Tennessee 37996, USA}

\author{Y. H. Liu}
\affiliation{Neutron Scattering Division, Oak Ridge National Laboratory, Oak Ridge, Tennessee 37831, USA}

\author{D. Parker}
\affiliation{Materials Science and Technology Division, Oak Ridge National Laboratory, Oak Ridge, Tennessee 37831, USA}

\author{Y. Wu}
\affiliation{Neutron Scattering Division, Oak Ridge National Laboratory, Oak Ridge, Tennessee 37831, USA}

\author{A. A. Aczel}
\affiliation{Neutron Scattering Division, Oak Ridge National Laboratory, Oak Ridge, Tennessee 37831, USA}
\affiliation{Department of Physics and Astronomy, University of Tennessee, Knoxville, Tennessee 37996, USA}

\author{M. Matsuda}
\affiliation{Neutron Scattering Division, Oak Ridge National Laboratory, Oak Ridge, Tennessee 37831, USA}

\author{M. A. McGuire}
\affiliation{Materials Science and Technology Division, Oak Ridge National Laboratory, Oak Ridge, Tennessee 37831, USA}

\author{B. C. Sales}
\affiliation{Materials Science and Technology Division, Oak Ridge National Laboratory, Oak Ridge, Tennessee 37831, USA}

\date{\today}

\begin{abstract}
MnBi$_4$Te$_{7}$ and MnBi$_6$Te$_{10}$ are two members with n=2 and 3 in the family of
MnBi$_{2n}$Te$_{3n+1}$ where the n=1 member, MnBi$_2$Te$_{4}$, has been intensively investigated as the first intrinsic antiferromagnetic topological insulator. Here we report the A-type antiferromagnetic order in these two compounds by measuring magnetic properties, electrical and thermal transport, specific heat, and single crystal neutron diffraction. Both compounds order into an A-type antiferromagnetic structure as does MnBi$_2$Te$_{4}$ with ferromagnetic planes coupled antiferromagnetically along the crystallographic \textit{c} axis.  While no evidence for any in-plane ordered moment is found for MnBi$_2$Te$_{4}$ or MnBi$_6$Te$_{10}$, weak reflections at half-L positions along the [0 0 L] direction are observed for MnBi$_4$Te$_{7}$ suggesting an in-plane ordered moment around 0.15$\mu_{B}$/Mn. The ordering temperature, T$_N$, is 13\,K for MnBi$_4$Te$_{7}$ and 11\,K for MnBi$_6$Te$_{10}$. The magnetic order is also manifested in the anisotropic magnetic properties. For both compounds, the interlayer coupling is weak and a spin flip transition occurs when a magnetic field of around 1.6\,kOe is applied along the \textit{c}-axis at 2\,K. As observed in MnBi$_2$Te$_4$, when cooling across T$_N$, no anomaly was observed in the temperature dependence of thermopower. On the other hand, critical scattering effects are observed in thermal conductivity although the effect is less pronounced than that in MnBi$_2$Te$_{4}$.

\end{abstract}

\maketitle

\section{Introduction}

\begin{figure*} \centering \includegraphics [width = \textwidth] {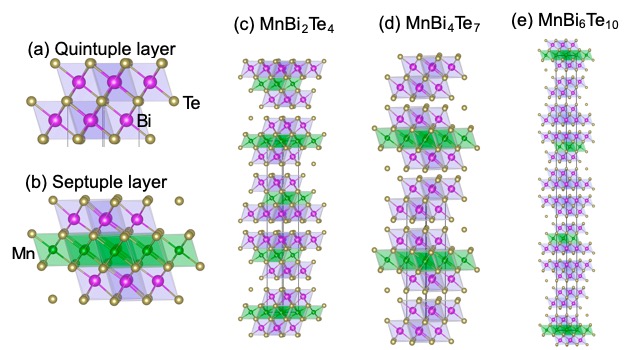}
\caption{(color online) The structure units: (a) The quintuple layer, (b) the septuple layer. The crystal structure of (c)  MnBi$_2$Te$_4$, (d) MnBi$_4$Te$_7$, and (e) MnBi$_6$Te$_{10}$.}
\label{Structure-1}
\end{figure*}

Antiferromagnetic topological insulators\cite{mong2010antiferromagnetic} can be an ideal materials playground for the study of various topological quantum phenomena, including the quantum anomalous Hall effect  and quantum axion electrodynamics. Recently, a layered cleavable transition metal chalcogenide, MnBi$_2$Te$_{4}$, was proposed\cite{otrokov2017highly, otrokov2018prediction} to be the first intrinsic antiferromagnetic topological insulator. As shown in Fig.\,\ref{Structure-1}, the rhombohedral crystal structure of MnBi$_2$Te$_4$ has an A-B-C stacking along the crystallographic \textit{c}-axis of the septuple layers which can be viewed as inserting one Mn-Te layer into a quintuple layer (see Fig.\,\ref{Structure-1}(b)). MnBi$_2$Te$_4$ is one member of a family of compounds with the chemical formula MnBi$_{2n}$Te$_{3n+1}$ and has n=1. Further spacing the magnetic septuple layers by the nonmagnetic quintuple layers along the \textit{c}-axis can lead to other MnBi$_{2n}$Te$_{3n+1}$ members with n$\geq$2. Previous experimental studies suggest the presence of compounds with n up to 5\cite{amiraslanov2018,souchay2019layered,aliev2019novel}. Figures\,\ref{Structure-1}(d,e) show the stacking of the septuple and quintuple layers along the \textit{c}-axis for MnBi$_4$Te$_{7}$ (n=2) and MnBi$_6$Te$_{10}$ (n=3).

Compared to MnBi$_2$Te$_4$, n$\geq$2 members have two unique features: (1) the magnetic septuple layers are further separated, which reduces the interlayer coupling, J$_c$. If the single ion anisotropy, D, is less affected, the reduced J$_c$ would significantly modify the magnetic response to the external field as observed in MnBi$_{2-x}$Sb$_x$Te$_4$\cite{yan2019evolution}; (2) The termination layer can be the magnetic septuple layers or the nonmagnetic quintuple layer. Thus quite different surface states can be expected. In addition, the termination quintuple layer might behave as a natural capping layer and affect the magnetic and electronic properties of the septuple layer right beneath it. Different stacking of the septuple and quintuple layers enriches the topological phenomena expected in MnBi$_{2n}$Te$_{3n+1}$ compounds\cite{sun2019rational}.

MnBi$_4$Te$_{7}$ and MnBi$_6$Te$_{10}$  have been theoretically proposed to be a possible materials platform for chiral Majorana fermions\cite{zhang2019plane}. The magnetic and transport properties, and the electronic structure of MnBi$_4$Te$_{7}$ have been experimentally investigated on single crystals by three different groups\cite{hu2020van,wu2019natural,vidal2019topological}. While all report an antiferromagnetic order below T$_N$=13\,K, Mn deficiency seems to introduce one extra magnetic order around 5\,K which further enriches the topological phase diagram\cite{vidal2019topological}. The anisotropic magnetic properties suggest the ordered moment is along the crystallographic \textit{c}-axis.  An A-type antiferromagnetic structure is likely while a neutron diffraction confirmation is still absent. The magnetic measurements on polycrystalline MnBi$_6$Te$_{10}$ suggest a ferromagnetic order with T$_C$=12\,K\cite{souchay2019layered}. An investigation of the intrinsic and anisotropic properties using single crystal samples is still needed.

In this work, we report the magnetic, transport, and thermodynamic properties of  MnBi$_4$Te$_{7}$  and MnBi$_6$Te$_{10}$ single crystals. We also report the magnetic structure determined by single crystal neutron diffraction. Magnetic order above 10\,K persists even though the \textit{c}-axis distance between interlayers of Mn approaches dozens of angstroms. Both compounds order into an A-type antiferromagnetic structure as does MnBi$_2$Te$_4$. The magnetic ordering temperature of 13\,K for MnBi$_4$Te$_{7}$ is slightly higher than 11\,K for MnBi$_6$Te$_{10}$. In both compounds, the interlayer coupling J$_c$ is significantly reduced by the increased spacing between the magnetic septuple layers and is much weaker than the single ion anisotropy. Therefore, as in MnSb$_2$Te$_{4}$, a small magnetic field of 1.6\,kOe along the \textit{c}-axis is enough to flip the moments at 2\,K.

\section{Experimental details}

MnBi$_4$Te$_7$ and MnBi$_6$Te$_{10}$ single crystals were grown out of a Bi-Te flux using the same starting materials and cooling process as for the growth of MnBi$_2$Te$_4$\cite{yan2019crystal}. According to the phase diagram developed by Aliev et al\cite{aliev2019novel}, MnBi$_4$Te$_7$ and MnBi$_6$Te$_{10}$ should crystallize at a lower temperature than that for MnBi$_2$Te$_4$. The crystallization of MnBi$_4$Te$_7$ and MnBi$_6$Te$_{10}$ using the same growth procedure for MnBi$_2$Te$_4$ might result from a natural temperature gradient of the box furnace. However, the flux can be well separated from crystals by decanting in a centrifuge at a temperature of 585$^\circ$C. While further study is needed to reveal the growth mechanism, we noticed that an overnight dwelling at 595$^{\circ}$C seems to favor the crystallization of MnBi$_4$Te$_7$ and MnBi$_6$Te$_{10}$ and the growths are reproducible.

\begin{figure} \centering \includegraphics [width = 0.47\textwidth] {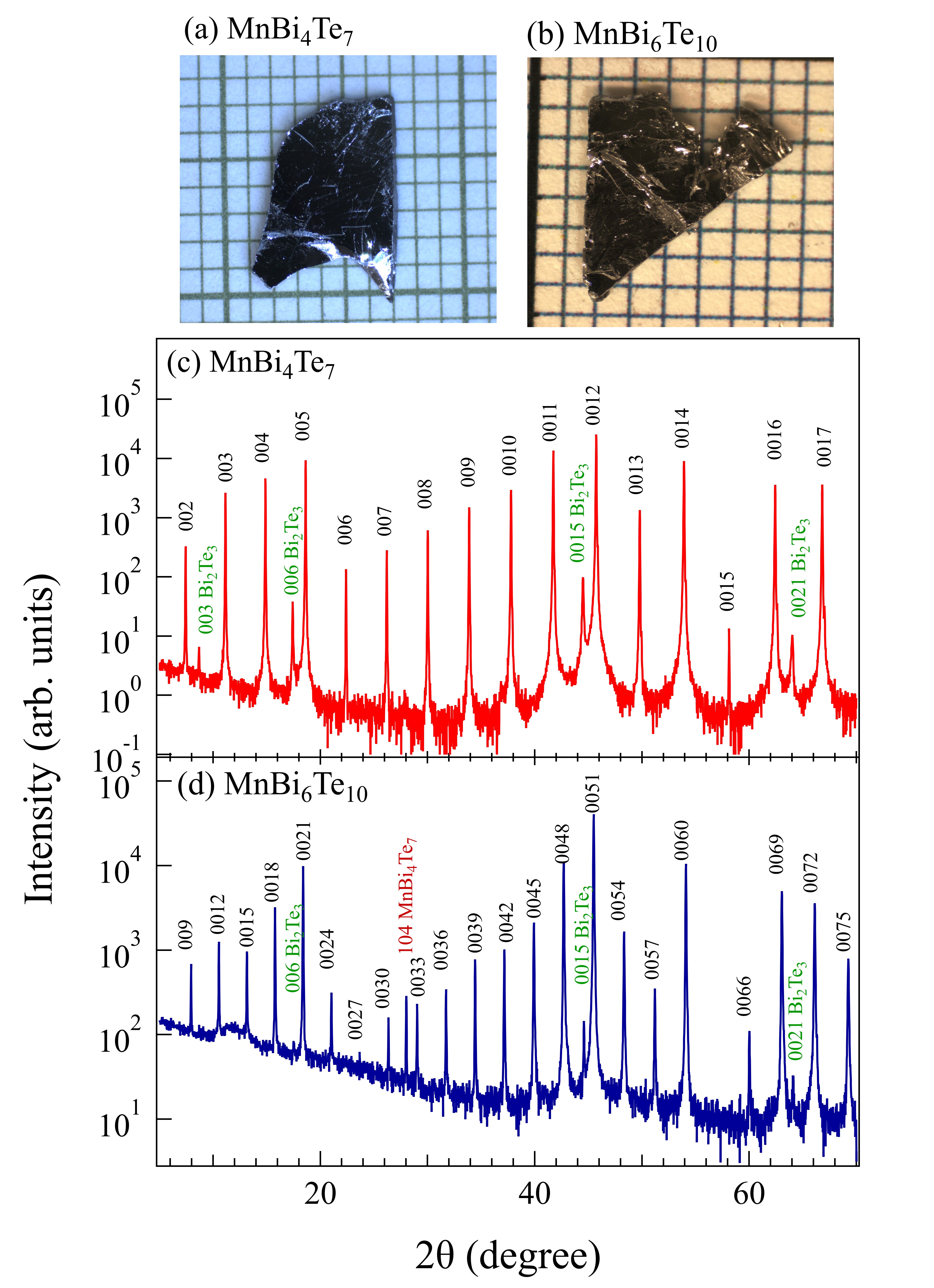}
\caption{(color online) The pictures of single crystals of (a) MnBi$_4$Te$_7$ and (b) MnBi$_6$Te$_{10}$ on a millimeter grid. The in-plane dimension is normally around several millimeters. The crystals tend to be rather thick  and some pieces can be over 1\,mm thick.  X-ray diffraction patterns of the surface of (c) MnBi$_4$Te$_7$ and (d) MnBi$_6$Te$_{10}$ single crystals. The observed ($00l$) reflections suggest that the \textit{c}-axis is perpendicular to the plane of the plates. The reflections are indexed with the structure proposed in Ref [\citenum{aliev2019novel}].  Weak reflections from Bi$_2$Te$_3$ in (c,d) and MnBi$_4$Te$_7$ in (d) are also indexed. Note log-scale for the intensity.}
\label{XRD-1}
\end{figure}

Figures\,\ref{XRD-1}(a) and (b) show the pictures of the plate-like crystals. It is interesting to note that MnBi$_4$Te$_7$ and MnBi$_6$Te$_{10}$ crystals grown in this manner tend to be rather thick (typically over 1\,mm). Elemental analysis on cleaved surfaces was performed to determine the elemental ratio using a Hitachi TM-3000 tabletop electron microscope equipped with a Bruker Quantax 70 energy dispersive x-ray system. Some small crystals were ground together with fine power of silicon\cite{mcguire2019chemical,yan2019evolution}  for x-ray powder diffraction measurements, which were performed  at room temperature using a PANalytical XPert Pro diffractometer with Cu-K$_{\alpha1}$ radiation.

 Magnetic properties in fields up to 70\,kOe were measured with a Quantum Design (QD) Magnetic Property Measurement System (MPMS). Magnetization data in fields up to 120\,kOe were collected using the ac option of a QD Physical Property Measurement System (PPMS). The temperature and field dependent electrical resistivity data were collected using a QD Physical Property Measurement System (PPMS). The temperature dependence of thermal conductivity and thermopower data were collected using the thermal transport option of PPMS on rectangular bars with the dimensions of 1.2\,mm\,$\times$\,0.75\,mm\,$\times$\,7\,mm cut from large plate-like crystals. Silver epoxy (H20E Epo-Tek) was utilized to provide mechanical and thermal contacts during the thermal transport measurements. The thermal conductivity measurement was performed with the heat flow in the \textit{ab}-plane.

Single crystal neutron diffraction experiments at temperatures down to 6.5\,K were carried out at the CORELLI spectrometer~\cite{ye2018implementation} at the Spallation Neutron Source (SNS), Oak Ridge National Laboratory.  The sizes of the two plate-like crystals used in the experiment are about $4 \times 5 \times 1~mm^3$ for MnBi$_4$Te$_7$ and $3 \times 4 \times 0.6$~mm$^3$ for MnBi$_6$Te$_{10}$, respectively. Both samples were mounted with the $c$-axis in the horizontal plane. Neutron-absorbing Cd was used to shield the sample holder to reduce the background scattering. Experiments were conducted by rotating the sample through a 180$^\circ$ range with a 2$^\circ$ step both below and above the magnetic transition temperatures to better isolate the magnetic Bragg scattering contributions. Order-parameter type data were collected as a function of temperature at a few fixed angles optimized for signals from particular magnetic peaks of interest. The Mantid package was used for data reduction including Lorentz and spectrum corrections~\cite{michels2016expanding}.  The integrated Bragg intensities were obtained from integration in the 3D reciprocal space and were corrected for background.  Possible magnetic structures were investigated by representation analysis using the SARAh program~\cite{wills2000new} and by the magnetic symmetry approach using the MAXMAGN program available at the Bilbao Crystallographic Server~\cite{perez2015symmetry}.  The magnetic structural refinements were carried out with the FullProf Suite program~\cite{rodriguez1993recent}.

Additional elastic neutron scattering and neutron diffraction experiments were performed at the triple axis spectrometers HB-1 and HB-1A and the four circle diffractometer HB-3A respectively, which are located at the High Flux Isotope Reactor (HFIR) of ORNL. All the measurements at HFIR are motivated to investigate whether the long-range ordered magnetic moments have a finite component in the $ab$ plane. The same pieces of MnBi$_4$Te$_7$ and MnBi$_6$Te$_{10}$ were used in all the neutron diffraction measurements. The MnBi$_2$Te$_4$ crystal used in this work has been well characterized as reported before \cite{yan2019crystal}. The $\theta$-2$\theta$ scans along the [0 0 L] direction were  performed at HB-1A for MnBi$_2$Te$_4$  and MnBi$_4$Te$_7$ and at HB-1 for MnBi$_6$Te$_{10}$. The same piece of MnBi$_4$Te$_7$ crystal was also measured at HB-3A in an effort of resolving the detailed magnetic structure. All three experiments used closed cycle refrigerators to achieve a base temperature comparable to the CORELLI experiment. The collimations for HB-1 and HB-1A were 40'-80'-80'-240' and 40'-40'-40'-80' respectively, while the incident neutron energy was 13.5 meV and 14.6 meV. Both triple axis instruments used pyrolytic graphite analysers for energy discrimination, leading to an energy resolution at the elastic line of $\sim$~1 meV (full-width half-maximum).

\begin{figure} \centering \includegraphics [width = 0.47\textwidth] {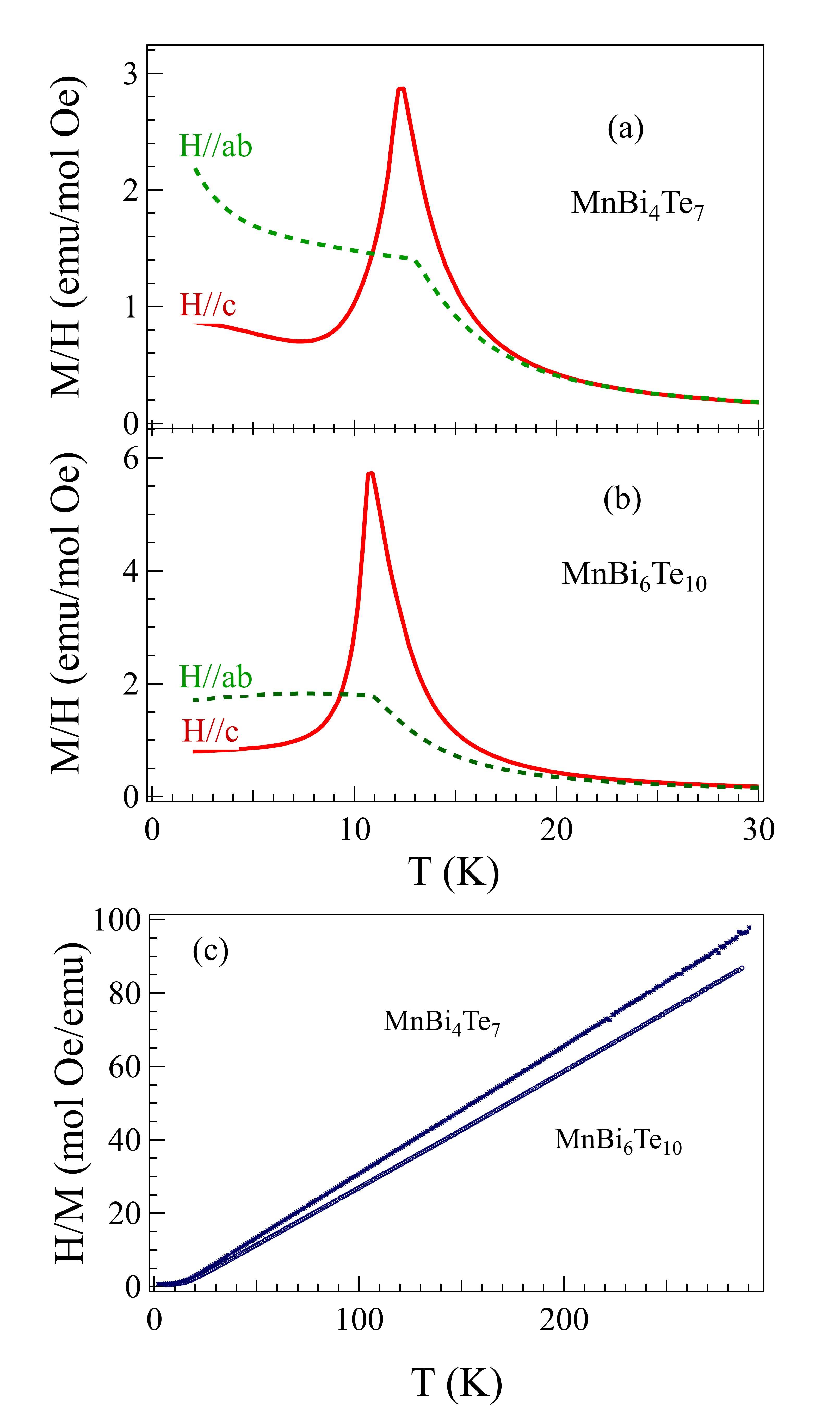}
\caption{(color online) Temperature dependence of magnetic susceptibility in an applied magnetic field of 250\,Oe perpendicular (H//\textit{ab})  and parallel (H//\textit{c}) to the crystallographic $c$ axis for (a) MnBi$_4$Te$_7$ and (b) MnBi$_6$Te$_{10}$. (c) Temperature dependence of reciprocal magnetic susceptibility, H/M, measured in a field of 10\,kOe applied perpendicular to the crystallographic \textit{c} axis.}
\label{chi}
\end{figure}

\section{Results}

\subsection{Elemental analysis and X-ray}
The elemental analysis was performed on flat surfaces cleaved in air. Multiple pieces from each batch were checked and no composition variation was observed. The measurements give a composition of Mn$_{6.8}$Bi$_{34.5}$Te$_{58.7}$ for the nominal MnBi$_4$Te$_7$. This suggests the MnBi$_4$Te$_7$ crystals are Bi rich but Mn deficient. The Mn site might host the extra Bi and can also have Mn vacancy. With this assumption, the chemical formula can be written as (Mn$_{0.81}$Bi$_{0.12}\Box_{0.07}$)Bi$_4$Te$_7$. This composition is similar to that reported by Vidal et al\cite{vidal2019topological}. As presented later, our MnBi$_4$Te$_7$ crystals also order antiferromagnetically at T$_N$=13\,K, which is consistent with previous reports\cite{hu2020van,wu2019natural,vidal2019topological}. The elemental analysis for the nominal MnBi$_6$Te$_{10}$ crystals gives a composition of Mn$_{4.4}$Bi$_{36.6}$Te$_{59.0}$, corresponding to (Mn$_{0.75}$Bi$_{0.2}\Box_{0.05}$)Bi$_6$Te$_{10}$ assuming the presence of Mn vacancies and extra Bi on the Mn site. For simplicity, we still label the crystals as MnBi$_4$Te$_7$ and MnBi$_6$Te$_{10}$ in the following text. The elemental analyses suggest that both MnBi$_4$Te$_7$ and MnBi$_6$Te$_{10}$ crystals grown out of Bi-Te flux tend to be Bi-rich but Mn-deficient. With the above nonstoichiometry and occupancy, both compounds are expected to be n-type, which is confirmed by the Hall measurements presented later. It should be noted that MnBi$_2$Te$_4$ crystals grown using the same procedure tend to be stoichiometric\cite{yan2019crystal}. Considering the growths of all three compounds start with the same starting materials, the Mn deficiency observed in MnBi$_4$Te$_7$ and MnBi$_6$Te$_{10}$  crystals indicates that a small percentage of Mn in Bi$_2$Te$_3$ flux can lower the melting temperature of the flux or the diffusivity of Mn in the melt is rather sensitive to temperature and plays an important role in selecting the precipitation. It should be noted that in the above analysis, we did not consider possible Mn ions at the Bi sites. The concentration of this substitutional defect might be quite low. In our neutron diffraction study, purposely putting a small fraction of Mn at the Bi sites can slightly improve the refinement. The concentration of this site mixing and its possible effects on the electronic properties deserve further study.

Figures\,\ref{XRD-1}(c) and (d) present the XRD patterns of the surface of as-grown single crystals along the \textit{c}-axis of MnBi$_4$Te$_7$ and MnBi$_6$Te$_{10}$. The $(00l)$ reflections can be indexed with the structure proposed in a previous report\cite{aliev2019novel}. It is interesting to note that a few weak reflections from Bi$_2$Te$_3$ are observed for both compounds and the (104) reflection from MnBi$_4$Te$_7$ is observed in the MnBi$_6$Te$_{10}$ pattern shown in Figure\,\ref{XRD-1}(d). The rather weak reflections from Bi$_2$Te$_3$ and MnBi$_4$Te$_7$ are also observed in the powder diffraction pattern (not shown) taken on pulverized MnBi$_6$Te$_{10}$ crystals. Once the same x-ray diffraction measurements were performed on cleaved surfaces, the weak reflections from Bi$_2$Te$_3$ can still be observed occassionally for both MnBi$_4$Te$_7$ and MnBi$_6$Te$_{10}$. This observation suggests that the impurities mainly stay on the surface of as-grown crystals as in MnBi$_2$Te$_4$\cite{yan2019crystal} although a small fraction of inclusions or intergrowths might also be likely. Considering the complex stacking along the \textit{c}-axis in MnBi$_{2n}$Te$_{3n+1}$ family, stacking faults are expected and the amount of stacking disorder might increase with increasing n. While a small fraction of stacking disorder may not significantly alter the bulk properties, it may affect the results in surface sensitive measurements such as angle resolved photoemission spectroscopy or scanning tunnelling microscopy, and may dominate the transport properties of thin flakes which happen to have stacking defects.

Since a satisfactory refinement of the powder diffraction pattern cannot be obtained, we estimated the volume fraction of MnBi$_4$Te$_7$ in the the MnBi$_6$Te$_{10}$ crystals to be around 3\% by simply comparing the intensity of the strongest peak of each phase. Using the structure model reported previously\cite{souchay2019layered,aliev2019novel}, the lattice parameters at room temperature are \textit{a}=4.3658\AA, \textit{c}=23.80\AA\,   for MnBi$_4$Te$_7$, and \textit{a}=4.374\AA, \textit{c}=101.93\AA\, for MnBi$_6$Te$_{10}$, which are in good agreement with previous reports\cite{souchay2019layered,aliev2019novel}.

\subsection{Magnetic properties}

Figure\,\ref{chi} shows the temperature dependence of the magnetic susceptibility measured in an applied magnetic field of 250\,Oe applied perpendicular (labelled as H//\textit{ab}) and parallel (H//\textit{c}) to the crystallographic \textit{c}-axis. MnBi$_4$Te$_7$ and MnBi$_6$Te$_{10}$ show similar anisotropic temperature dependent magnetic susceptibility. The anisotropic temperature dependence suggests an antiferromagnetic order and the ordering temperature is T$_N$=13\,K for MnBi$_4$Te$_7$  and T$_N$=11\,K for MnBi$_6$Te$_{10}$. T$_N$=13\,K for MnBi$_4$Te$_7$ agrees with previous reports\cite{hu2020van,wu2019natural,vidal2019topological}. For both compounds, the magnetic susceptibility curves measured with H//\textit{ab} and H//\textit{c} in the temperature range 50$\leq$T$\leq$300\,K can be described by the Curie-Weiss law, $\chi=C/(T-\theta$), where C is the Curie constant and $\theta$ is the Weiss temperature. Figure\,\ref{chi}(c) shows the temperature dependence of reciprocal magnetic susceptibility, H/M, measured in a field of 10\,kOe applied perpendicular to the crystallographic \textit{c} axis. The linear fitting of the high temperature susceptibility data of MnBi$_6$Te$_{10}$ obtains a Weiss temperature of 12\,K and an effective moment of 5.0(1)\,$\mu_B$/Mn. A similar fitting for MnBi$_4$Te$_7$  obtains a Weiss temperature of 11\,K and an effective moment of 5.2(1)\,$\mu_B$/Mn. The positive Weiss temperature suggests ferromagnetic correlations in the paramagnetic state despite the low temperature antiferromagnetic order. It is worth mentioning that all n=1,2,and 3 members have a positive Weiss temperature even though they all order antiferromagnetically at low temperatures. The Curie-Weiss-like tail below $\approx$8\,K in the temperature dependence of magnetic susceptibility of MnBi$_4$Te$_7$ is different from a nearly temperature independent susceptibility for MnBi$_6$Te$_{10}$. We carefully cleaned the crystal surface and edges to reduce the possible contribution from magnetic impurities\cite{yan2019crystal}. However, the Curie-Weiss-like tail is always observed. This Curie-Weiss-like tail is also observed in two previous reports\cite{hu2020van,wu2019natural}. Whether the low-temperature Curie-Weiss-like feature is an intrinsic property of MnBi$_4$Te$_7$ deserves further study. We  noticed that the composition of our  MnBi$_4$Te$_7$ single crystals is similar to that reported by Vidal et al\cite{vidal2019topological}. However, the magnetic properties are different especially below 5\,K.  These discrepancies may arise from different concentration and/or distribution of Mn/Bi site mixing and Mn vacancies.

\begin{figure} \centering \includegraphics [width = 0.47\textwidth] {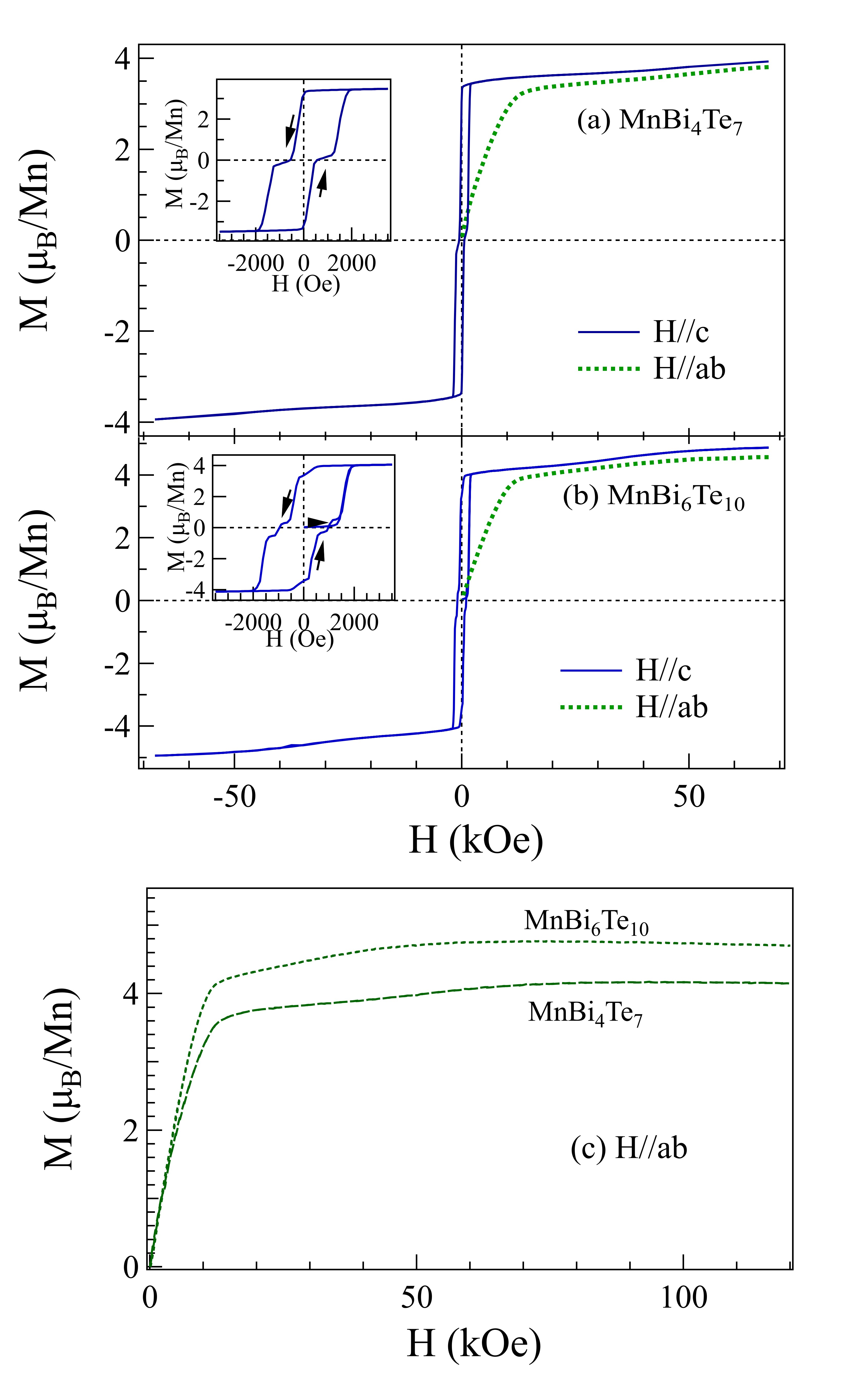}
\caption{(color online) Field dependence of magnetization at 2\,K in magnetic fields up to 70\,kOe for (a) MnBi$_4$Te$_7$ and (b) MnBi$_6$Te$_{10}$. For H//\textit{ab}, the measurements were done with decreasing field from 70\,kOe to 0\,Oe. For H//\textit{c}, the magnetization data for MnBi$_4$Te$_7$ were collected with decreasing field from 70\,kOe to -70\,Oe, and then with increasing field from -70\,kOe to 70\,Oe. For MnBi$_6$Te$_{10}$, the virgin curve was also measured. The insets highlight the details around zero field for the magnetization curves measured with H//\textit{c}. The arrows in the insets show the sequence of field changing. (c) Field dependence of magnetization at 2\,K in magnetic fields up to 120\,kOe. The data were collected with H//\textit{ab}.}
\label{MH}
\end{figure}

Figure\,\ref{MH} shows the magnetization, M(H), curves at 2\,K in fields up to 70\,kOe applied both along and perpendicular to the crystallographic \textit{c} axis. For both compounds, a spin flip transition occurs around a field of 1.6\,kOe when the field is parallel to the \textit{c} axis (H//\textit{c}). Once the field is applied perpendicular to the \textit{c} axis (H//\textit{ab}), the magnetization increases with field and then tends to saturate above 10\,kOe. The field dependence of magnetization resembles that of MnSb$_2$Te$_4$\cite{yan2019evolution}. As discussed later, this is because the interlayer coupling is significantly reduced by the increased spacing between the magnetic layers while the single ion anisotropy remains more or less unchanged compared to that in MnBi$_2$Te$_4$. There are two additional features of interest in 
in Fig.\,\ref{MH}: (1) Between 40-50\,kOe, the magnetization shows a 10\% increase and no hysteresis was observed in this field range. It is interesting to note that this 10\% enhancement occurs in both compounds and for both H//\textit{c} and H//\textit{ab}. If it originates from an impurity phase, the isotropic behavior might signal a polycrystalline impurity, such as Mn-doped Bi$_2$Te$_3$ flux or the presence of defect spins (such as Mn at Bi sites) that are anti-aligned to the Mn spins. (2) The other feature is the steps in the M(H) curves around zero magnetization which are highlighted in the Fig.\,\ref{MH} insets. The origin of this anomaly is unknown and it might be correlated with the 10\% enhancement at high fields. The origin of both features deserves further study. 

The field dependence of the magnetization was further measured in magnetic fields up to 120\,kOe at 2\,K. Figure\,\ref{MH}(c) shows the data collected with H//\textit{ab}. Above 60-70\,kOe, the magnetization saturates and no other field-induced transitions are observed up to 120\,kOe. At 2\,K, the saturation moment is 4.16\,$\mu_B$/Mn for MnBi$_4$Te$_7$ and 4.75\,$\mu_B$/Mn for MnBi$_6$Te$_{10}$. As listed in Table III, these saturation moments  are larger than that of MnBi$_2$Te$_4$.

\subsection{Transport properties}

\begin{figure} \centering \includegraphics [width = 0.47\textwidth] {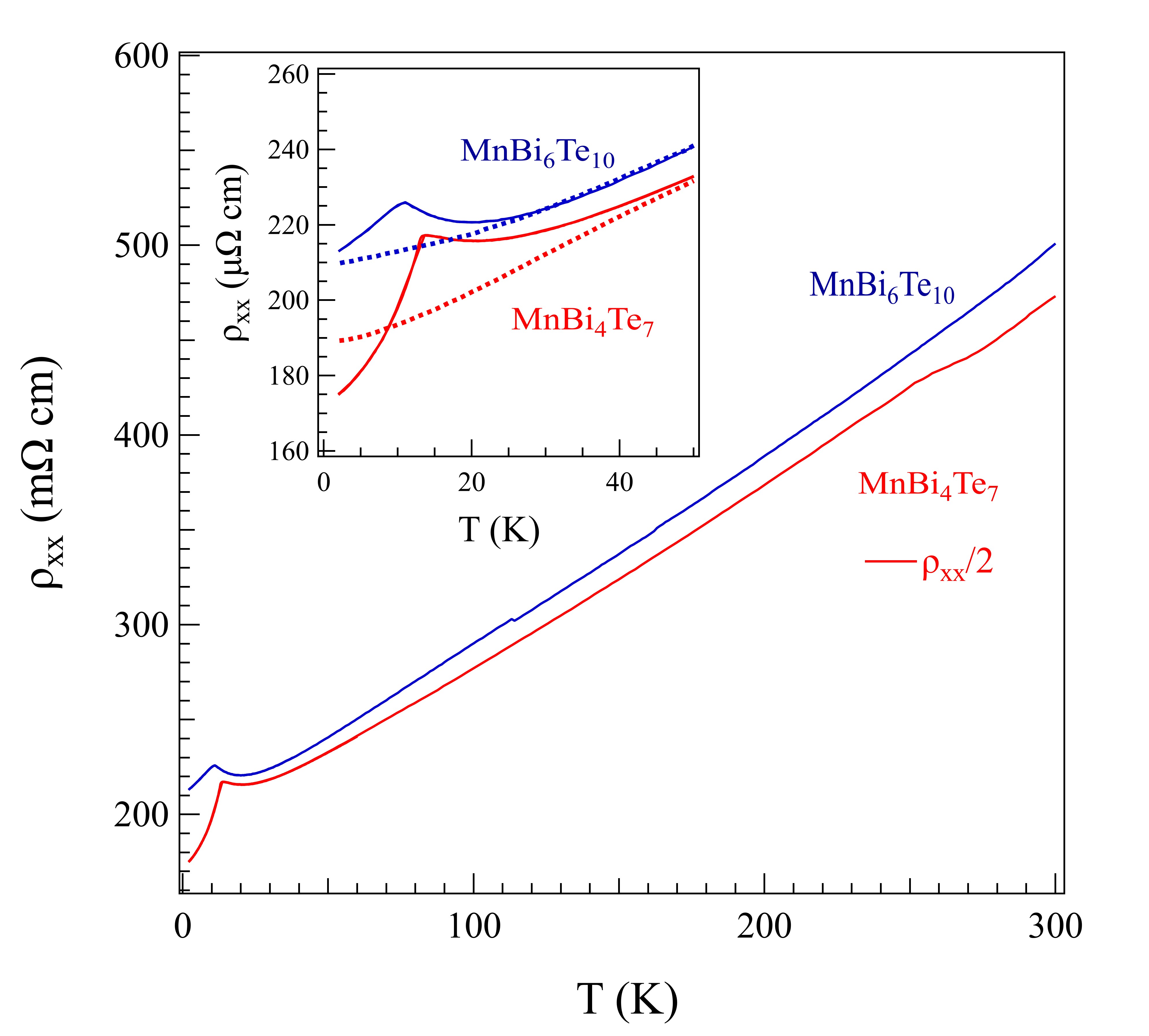}
\caption{(color online) Temperature dependence of in-plane electrical resistivity with the electrical current flowing in the \textit{ab}-plane. The inset highlights the details at the low temperatures as well as the effects of a magnetic field of 80\,kOe applied along the crystallographic \textit{c}-axis. Solid curves: zero field. Dashed curves: 80\,kOe.}
\label{Rxx}
\end{figure}

Figure\,\ref{Rxx} shows the temperature and field dependence of the in-plane electrical resistivity measured in the temperature range 2\,K$\leq$T$\leq$300\,K with the electrical current flowing in the \textit{ab}-plane. Both compounds show a metallic conducting behavior and a cusp around T$_N$. Compared to MnBi$_2$Te$_4$, both compounds are more conductive. The Fig.\,\ref{Rxx} inset highlights the details around T$_N$ and also the effect of a magnetic field of 80\,kOe applied along the crystallographic \textit{c}-axis. Both the temperature and field dependence of both compounds resembles that of MnBi$_2$Te$_{4}$.

\begin{figure} \centering \includegraphics [width = 0.47\textwidth] {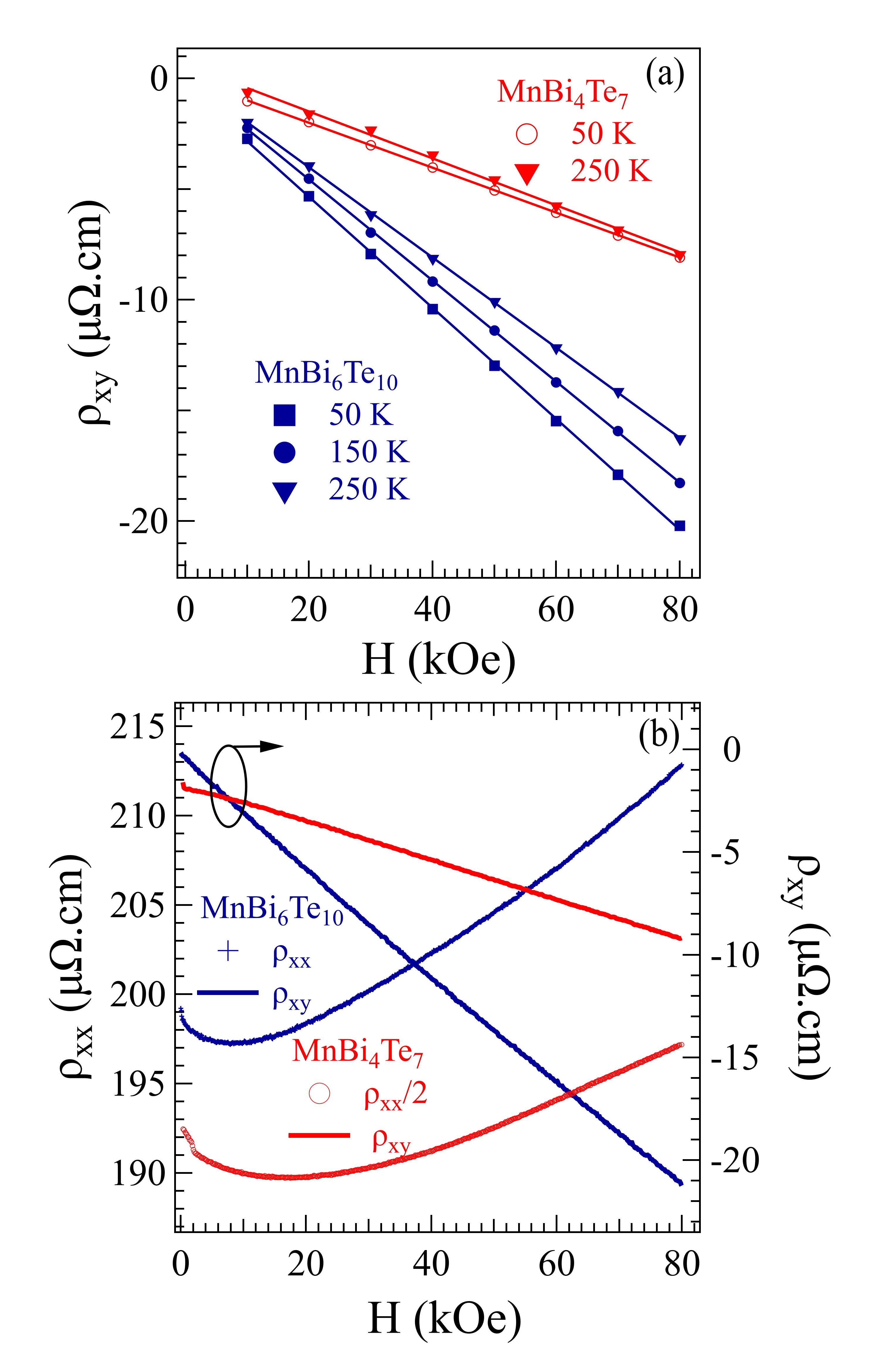}
\caption{(color online) (a) Hall resistivity versus applied magnetic field for temperatures between 50 to 300\,K with the electrical current in the \textit{ab}-plane and the magnetic field along the \textit{c}-axis. The solid lines highlight the linear field dependence. (b) $\rho_{xx}$ and $\rho_{xy}$ as a function of field at 2\,K. }
\label{RxxRxy-1}
\end{figure}

Figure\,\ref{RxxRxy-1}(a) shows the Hall resistivity at selected temperatures. A linear field dependence is observed at all temperatures for both compounds. The Hall coefficient shows little temperature dependence below room temperature. With the assumption that a single band dominates the Hall signal, the room temperature coefficient gives an electron carrier density of  6.0$\times$10$^{20}$cm$^{-3}$ and 3.5$\times$10$^{20}$cm$^{-3}$ for MnBi$_4$Te$_7$ and MnBi$_6$Te$_{10}$, respectively. Figure\,\ref{RxxRxy-1}(b) shows the field dependence of in-plane $\rho_{xx}$ and $\rho_{xy}$ at 2\,K in magnetic fields up to 80\,kOe applied along the \textit{c}-axis. While a linear field dependence of $\rho_{xy}$ is observed for both compounds, $\rho_{xx}$ decreases with increasing field reaching a minimum around 20\,kOe and 10\,kOe for MnBi$_4$Te$_7$ and MnBi$_6$Te$_{10}$, respectively, then increases with increasing field. The slight increase of $\rho_{xx}$ at high fields is similar to that of MnSb$_2$Te$_4$\cite{yan2019evolution} and is typical for a metal.

\begin{figure} \centering \includegraphics [width = 0.47\textwidth] {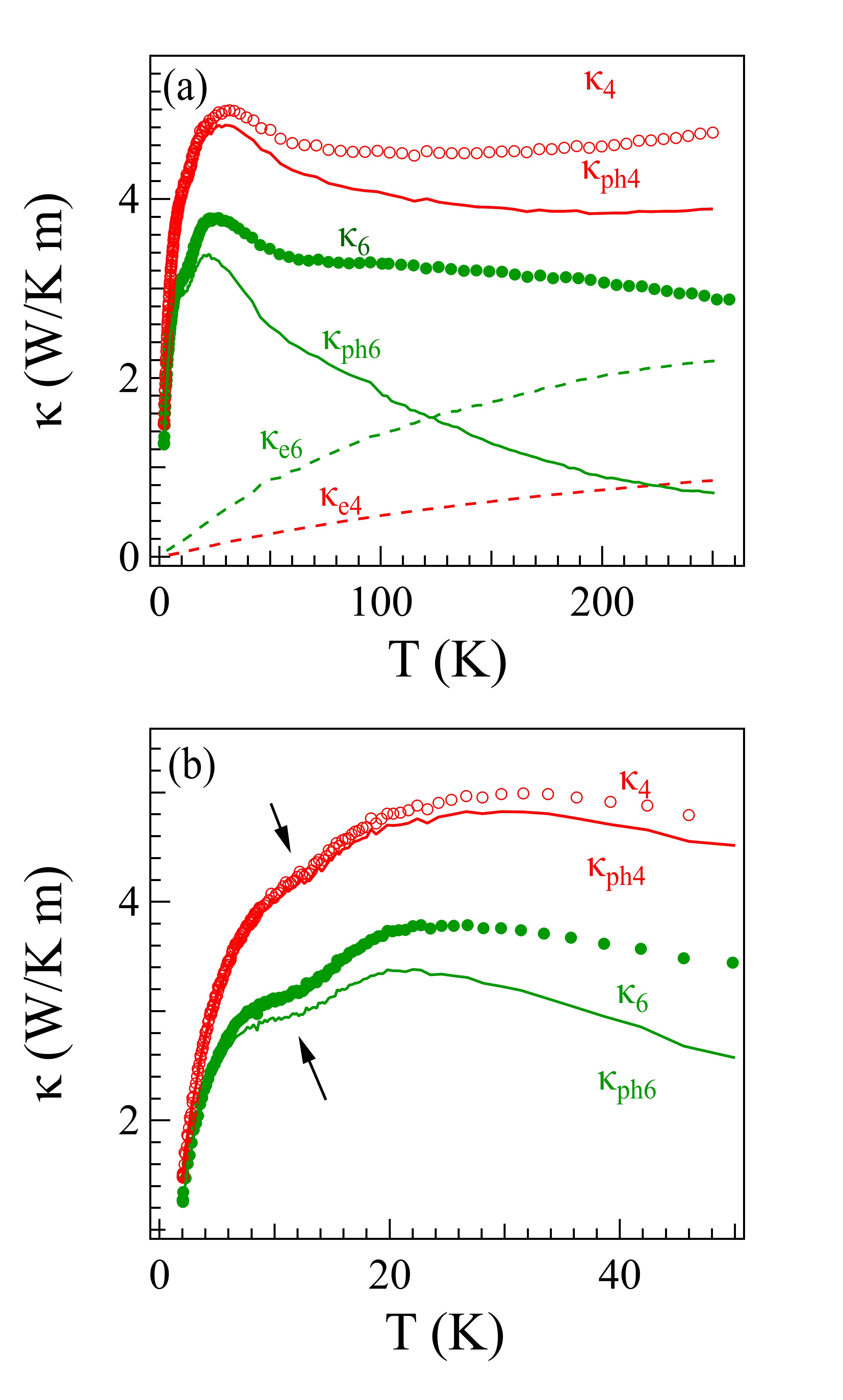}
\caption{(color online) (a) Temperature dependence of in-plane thermal conductivity measured with the heat current flowing in the \textit{ab}-plane in the temperature range 2\,K$\leq$T$\leq$250\,K. The electronic thermal conductivity, $\kappa_e$, was estimated from the electrical resistivity data using the Wiedemann-Franz law. The lattice thermal conductivity, $\kappa_{ph}$, was obtained by subtracting $\kappa_e$ from the total thermal conductivity. (b)  Details of the in-plane thermal conductivity around T$_N$. $\kappa_{4}$ is the thermal conductivity for MnBi$_4$Te$_7$.  $\kappa_{6}$ is the thermal conductivity for MnBi$_6$Te$_{10}$.}
\label{kappa}
\end{figure}

Figure \ref{kappa}(a) shows the temperature dependence of the in-plane thermal conductivity, $\kappa$(T), measured with the heat current flowing in the \textit{ab}-plane in the temperature range 2\,K$\leq$T$\leq$250\,K. $\kappa$(T) curves for both compounds show a maximum around 30\,K and little temperature dependence above 50\,K, similar to that of MnBi$_2$Te$_4$.  Around 250\,K, MnBi$_4$Te$_7$ has an in-plane thermal conductivity of $\approx$4.8\,W/K m, which is about 50\% larger than that of  MnBi$_6$Te$_{10}$  or MnBi$_2$Te$_4$. Figure \ref{kappa}(a) also shows the electronic thermal conductivity estimated following Wiedemann-Franz law and the phonon thermal conductivity obtained by subtracting the electronic thermal conductivity from the total thermal conductivity. While the phonon thermal conductivity dominates in MnBi$_2$Te$_4$ \cite{yan2019crystal} and MnBi$_4$Te$_7$, the electronic thermal conductivity in MnBi$_6$Te$_{10}$  dominates above $\approx$120\,K. Figure \ref{kappa}(b) highlights the details of $\kappa$(T) below 50\,K. Around T$_N$, both $\kappa$(T) curves show a dip-like feature, a typical feature for the critical scattering. However, the dip-like feature is less prominent than that in MnBi$_2$Te$_4$.

\begin{figure} \centering \includegraphics [width = 0.47\textwidth] {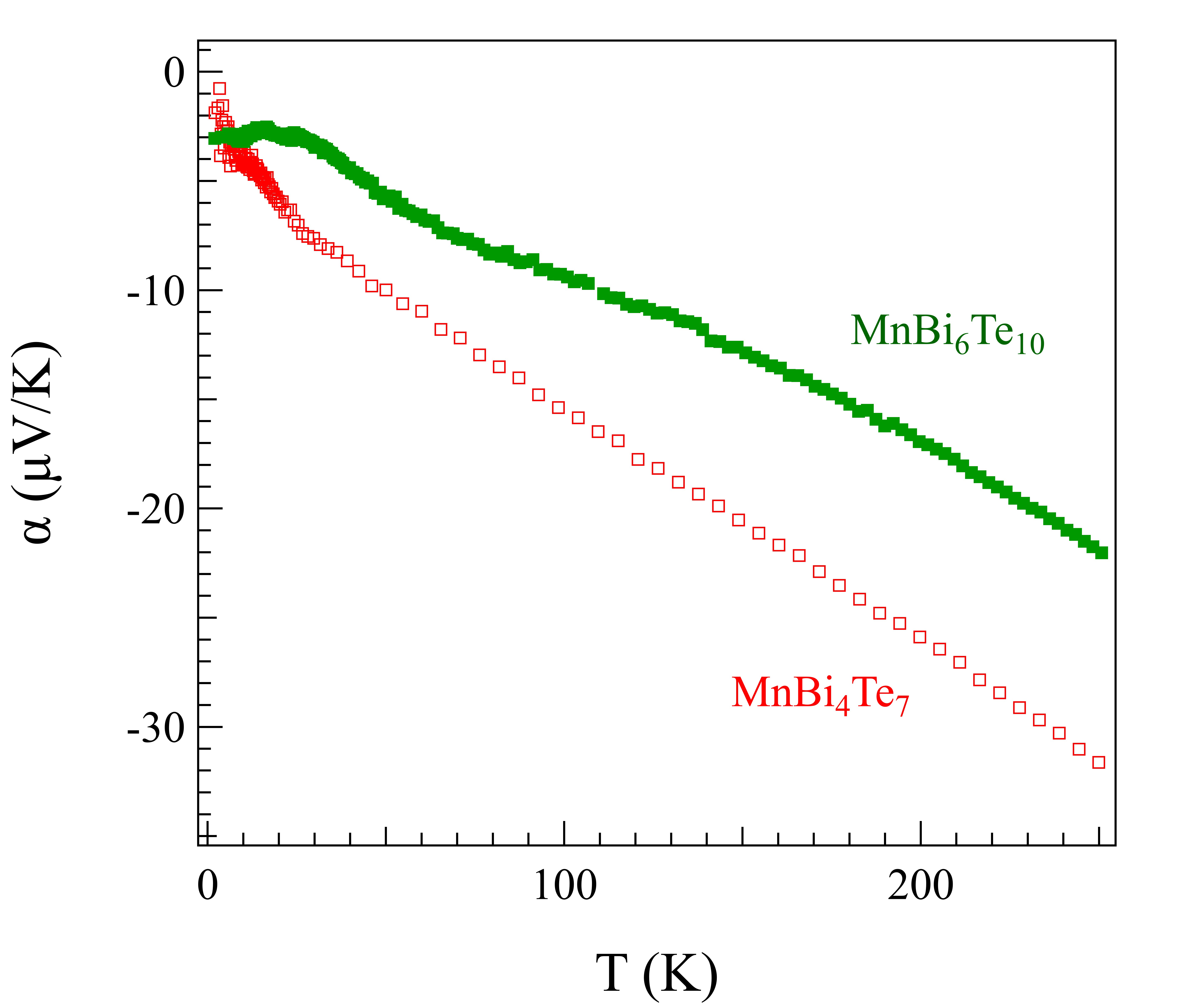}
\caption{(color online) Temperature dependence of thermopower. }
\label{Seeb}
\end{figure}

Figure\,\ref{Seeb} shows the temperature dependence of thermopower, $\alpha$(T), below room temperature. At room temperature, $\alpha$(T) has a value of -38$\mu$V/K and -26$\mu$V/K for MnBi$_4$Te$_{7}$ and MnBi$_6$Te$_{10}$, respectively. The negative sign of $\alpha$(T) signals electron dominated charge transport, consistent with the Hall data. The absolute value of $\alpha$(T) decreases upon cooling from room temperature to 2\,K. The nearly linear temperature dependence above 50\,K corresponds to the characteristic diffusion thermopower of a metal. Similar to that in MnBi$_2$Te$_{4}$, no response of $\alpha$(T) to the magnetic order was observed around T$_N$.

\subsection{Specific heat}
\begin{figure} \centering \includegraphics [width = 0.47\textwidth] {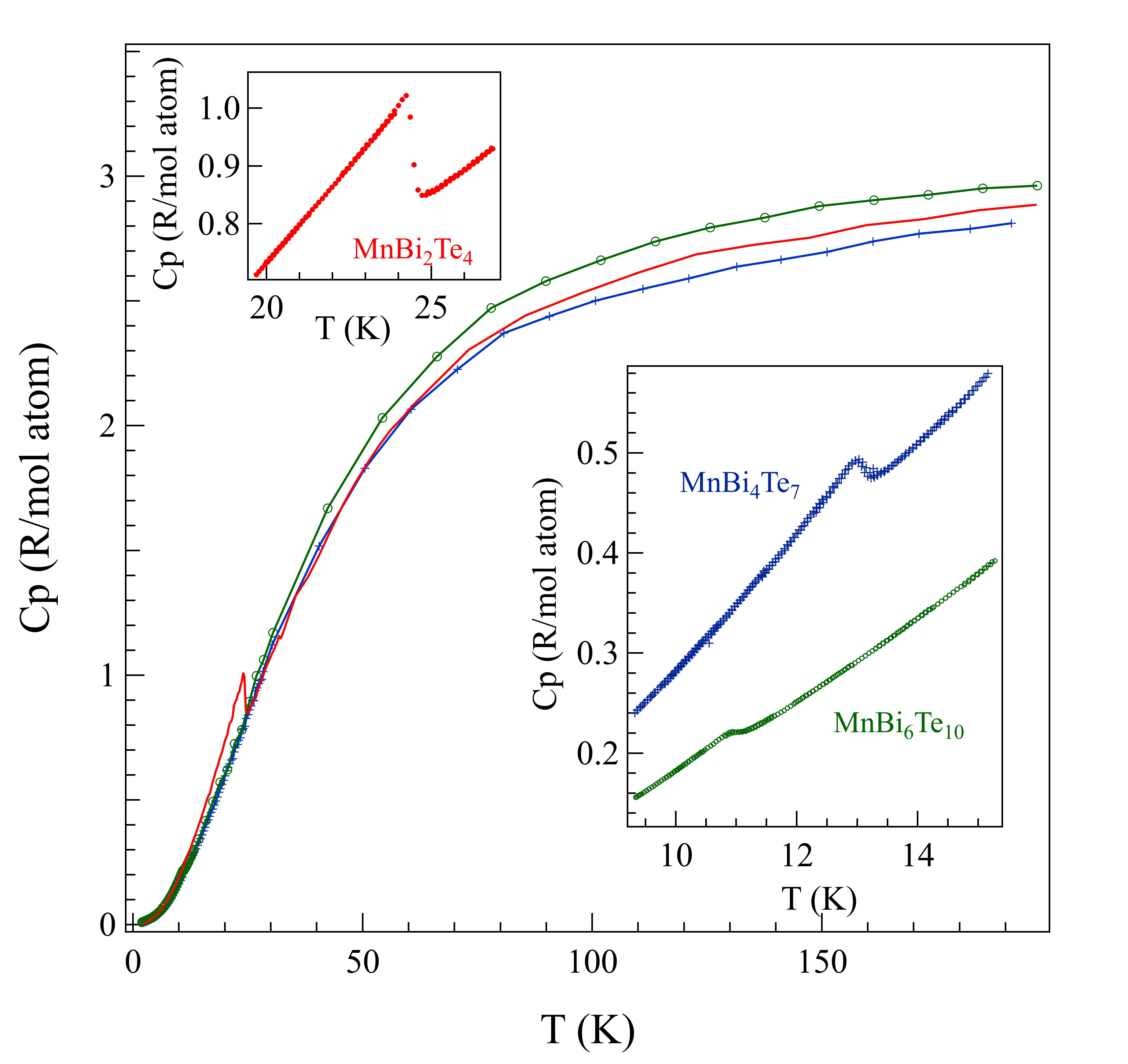}
\caption{(color online) Temperature dependence of specific heat below 200\,K. The insets highlight the details around T$_N$. }
\label{Cp-1}
\end{figure}

Figure\,\ref{Cp-1} shows the temperature dependence of specific heat below 200\,K for MnBi$_4$Te$_{7}$ and MnBi$_6$Te$_{10}$. The data for MnBi$_2$Te$_{4}$ is also plotted for comparison. The specific heat data approach the Dulong-Petit limit at high temperatures. As highlighted in the insets of the figure, there is a weak lambda-type anomaly around T$_N$ for each composition and the magnitude of the anomaly becomes smaller with increasing n. The weak anomaly around T$_N$ might result from the entropy release by the two dimensional (2D) magnetic correlations existing at higher temperatures as reported in other 2D cleavable magnets\cite{mcguire2015coupling,casto2015strong}. The low temperature specific heat data in the range 10$\leq T^2 \leq$60\,K$^2$ follows the standard power law, C$_P$=$\gamma T+\beta T^3$, where $\gamma$ is the Sommerfeld electronic specific heat coefficient and $\beta$ is the coefficient of the Debye T$^3$ lattice heat capacity at low temperatures. The fitting of the low temperature specific heat data gives the $\gamma$ value of 0.18, 0.34, and 0.67\,J/mol K$^2$ for MnBi$_2$Te$_{4}$, MnBi$_4$Te$_{7}$, and MnBi$_6$Te$_{10}$, respectively. The Debye temperature, $\theta_D$, can be estimated from the  $\beta$ coefficient using the equation  $\theta_D$=(12$\pi^4N_Ak_Bn/5\beta)^{1/3}$, where n is the number of atoms per formula unit, N$_A$ is Avogadro’s constant and k$_B$ is Boltzmann’s constant. All three compounds have a similar $\theta_D$, which is 110, 120, and 120\,K for MnBi$_2$Te$_{4}$, MnBi$_4$Te$_{7}$, and MnBi$_6$Te$_{10}$, respectively.

\subsection{Neutron single crystal diffraction}
\subsubsection{Magnetic Order and Critical Behavior}

\begin{figure}[tb]
	\centering
		\includegraphics[width=0.5\textwidth]{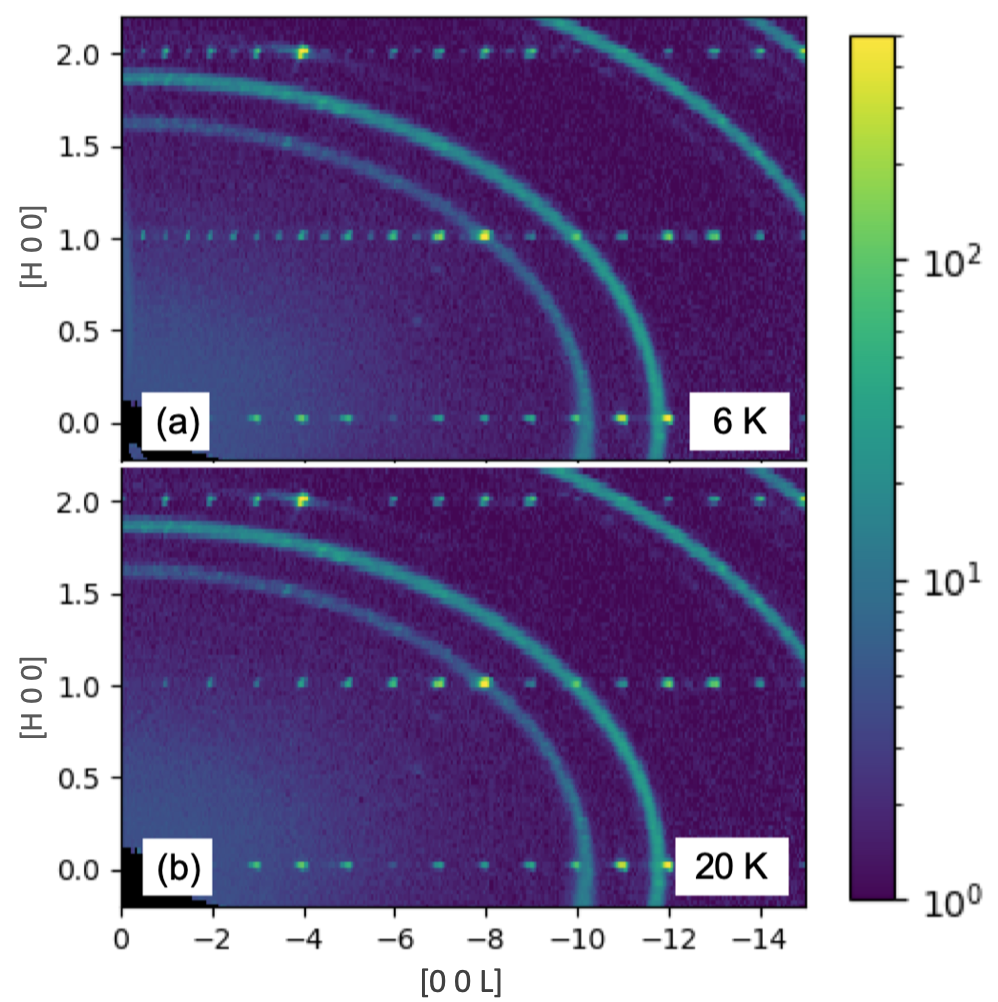}
		\caption{ \label{Fig:Mn147mesh} (Color online) 2D slice in the H0L plane of the neutron diffraction data from MnBi$_4$Te$_7$ collected at CORELLI at 6~K and 20~K, respectively. The Al powder rings in this figure arise from the sample environment and the sample mount.}
\end{figure}

\begin{figure}[tb]
	\centering
		\includegraphics[width=0.5\textwidth]{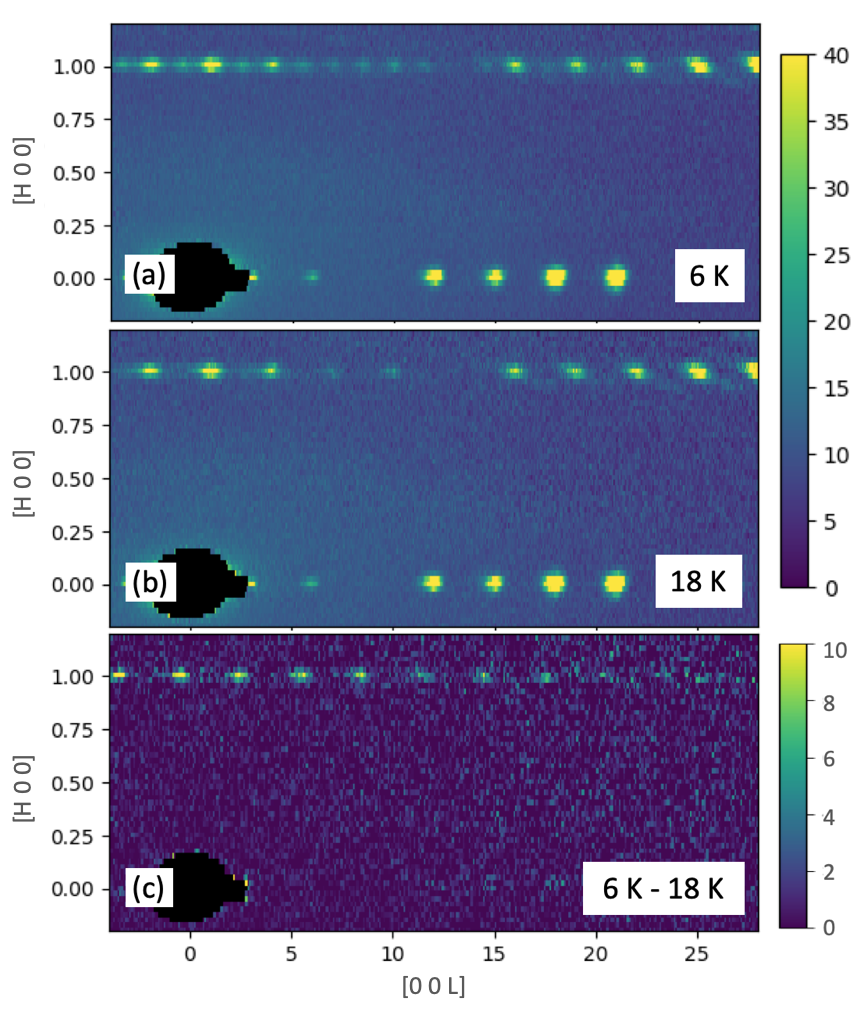}
		\caption{ \label{Fig:Mn1610mesh} (Color online)  2D slice in the H0L plane of the neutron diffraction data from MnBi$_6$Te$_{10}$ collected at CORELLI at 6~K (a) and 18~K (b), respectively, and (c) the difference between these two datasets. There is a notable signal at the magnetic Bragg peak positions from the tails of the nuclear Bragg peaks in the L direction. By taking the difference between 6 K and 18 K, this contribution can be largely removed, and the magnetic Bragg peaks are much more visible as seen in (c). Note that panel (c) uses a different intensity scale than (a) and (b).}
\end{figure}

Figure~\ref{Fig:Mn147mesh} shows the 2D slice in the H0L plane of the single-crystal neutron diffraction data from MnBi$_4$Te$_7$, collected at 6~K and 20~K, respectively. Comparing to the 20~K data, there are additional Bragg peaks at half-integer-L positions in the low Q regions at 6~K. These half-integer-L Bragg peaks become less visible as L increases, indicating that they originate from magnetic scattering.  The locations of these peaks agree with a magnetic wavevector of (0 0 0.5), suggesting that the magnetic unit cell is twice as large as the chemical unit cell along the $c$-axis. Note that the data in Fig.\,\ref{Fig:Mn147mesh} do not show any half-integer-L Bragg reflections along the [0 0 L] direction. However, as presented later, there are very weak but notable half-Bragg peaks along [0 0 L] for MnBi$_4$Te$_7$, which can only been seen from  high signal-to-noise ratio scans performed using a triple-axis spectrometor. Due to the vectorial nature of the neutron-magnetic moment interaction~\cite{halpern1939magnetic}, the extremely weak half-Bragg peaks along [0 0 L] suggests that the major component of the ordered magnetic moment is along the $c$-axis with a small in-plane component, consistent with the anisotropic magnetic properties.

Figures~\ref{Fig:Mn1610mesh}(a) and~\ref{Fig:Mn1610mesh}(b) show the similar 2D slice plots from the MnBi$_6$Te$_{10}$ datasets collected at 6 K and 18 K, respectively. New features along L in the [1 0 L] direction show up at 6~K when cooling down from 18~K. However, as seen from the 18~K dataset, the nuclear Bragg peaks have very long tails in the L direction. To better isolate the new scattering features, the difference between the two datasets are shown in Fig.~\ref{Fig:Mn1610mesh}(c). It becomes clear that the new features are Bragg-peak-like and located at positions of (1 0 3N-0.5), where N is an integer. The locations of these new peaks agree with a magnetic wavevector of (0 0 1.5).  There are no detectable half-integer-L Bragg peaks along the [0 0 L] direction. This is also confirmed by the measurements using the triple-axis spectrometers. Therefore the ordered magnetic moment is along the $c$-axis for MnBi$_6$Te$_{10}$.

\begin{figure}[tb]
	\centering
		\includegraphics[width=0.5\textwidth]{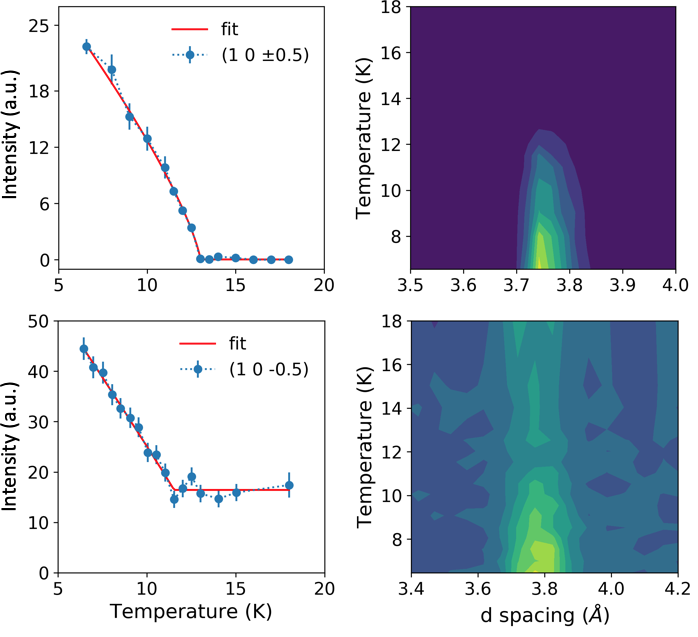}
		\caption{ \label{Fig:OPs} (Color online) Temperature dependence of peak intensities for selected low-Q magnetic peaks for  MnBi$_4$Te$_7$ (top) and MnBi$_6$Te$_{10}$ (bottom), respectively. The right two panels show the temperature dependence of the peak profiles. Note that there is a high background at the magnetic Bragg peak position for MnBi$_6$Te$_{10}$. The left two show the integrated peak intensities, where the dots sitting on dashed lines are the experimental data and the red solid lines are the best fits to the scaling equation of state.}
\end{figure}

Figure~\ref{Fig:OPs} shows the temperature dependence of some selected low $Q$ magnetic Bragg peaks. The 2D plots show that the high temperature background is clean for MnBi$_4$Te$_7$, but high for MnBi$_6$Te$_{10}$. As discussed above, the high background for MnBi$_6$Te$_{10}$ is from the tails of the adjacent nuclear Bragg Peaks. This has some effects on the peak intensity integration for the magnetic structure determination, which will be discussed later. Here we assume that the background is temperature independent and fit the temperature dependence of the integrated magnetic peak intensities using the scaling equation of state, $I (T) = I_{0} \times (1-T/T_N)^{2\beta} + I_{bac}$, where $\beta$ is the critical exponent, $T_N$ is the magnetic ordering temperature, and $I_{bac}$ is a constant. The magnetic Bragg peak intensity is proportional to the square of the spontaneous staggered magnetic moment; therefore, a factor of 2 is used before the critical exponent in the equation. The magnetic ordering temperatures from the best fits are $13.01~\pm~0.01$~K and $11.53~\pm~0.05$~K for MnBi$_4$Te$_7$ and MnBi$_6$Te$_{10}$, respectively, which are in a good agreement with their values obtained from the magnetic and transport measurements.  The effective critical exponents are $\beta = 0.39~\pm~0.01$ and  $\beta = 0.46~\pm~0.04$ for MnBi$_4$Te$_7$ and MnBi$_6$Te$_{10}$, respectively, in the studied temperature range.

We also examined the diffraction patterns for possible diffuse scattering to look for evidence of stacking disorder/faults in this family of materials. Clear rod-like diffuse scattering features are observed for MnBi$_6$Te$_{10}$ but are barely visible above the background for  MnBi$_4$Te$_7$. This suggests that the MnBi$_6$Te$_{10}$ crystals consist of more stacking faults. Detailed investigations of the stacking order in these materials are in progress.

\begin{figure*} \centering \includegraphics [width = \textwidth] {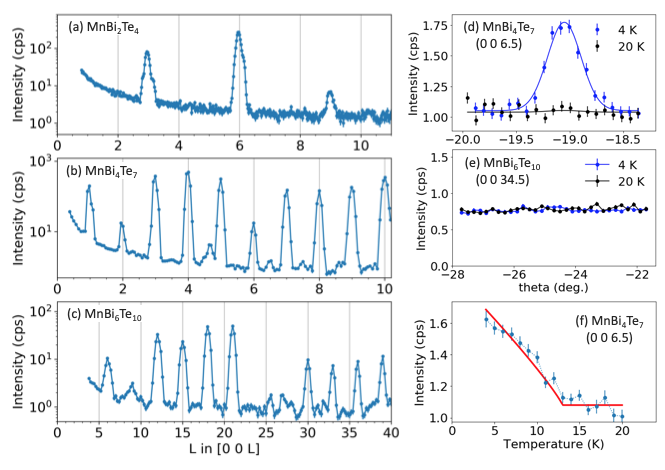}
\caption{(Color online) (a)-(c) $\theta-2\theta$ scans along [0 0 L]  taken at 4 K for MnBi$_2$Te$_{4}$, MnBi$_4$Te$_7$ and MnBi$_6$Te$_{10}$, respectively. There are very weak half-integer Bragg peaks only in MnBi$_4$Te$_7$, whose intensities are about 400 times weaker than other magnetic Bragg peaks.  (d) and (e): The rocking curves taken at half-integer Bragg peaks along [0 0 L] for (d) MnBi$_4$Te$_7$, (0 0 6.5) and (e) MnBi$_6$Te$_{10}$, (0 0 34.5),  collected at 4 K and 20 K. The signal shows a strong temperature dependence for MnBi$_4$Te$_7$, but is essentially featureless for MnBi$_6$Te$_{10}$. (f) Temperature dependence of the intensity of (0 0 6.5) for  MnBi$_4$Te$_7$.}
\label{Extra-1}
\end{figure*}

\subsubsection{In-plane Ordered Moments?}

To investigate whether the magnetic moments have a finite ordered component in the $ab$ plane, we performed elastic neutron scattering experiments using the two triple axis spectrometers HB-1 and HB-1A at the HFIR. We collected the triple axis data on MnBi$_2$Te$_4$, MnBi$_4$Te$_7$, and MnBi$_6$Te$_{10}$; $\theta-2\theta$ scans along the [0 0 L] direction are shown in Fig.\,\ref{Extra-1}(a)-(c). As illustrated in Fig. \,\ref{Extra-1}(b), there are weak peaks at half-L positions for MnBi$_4$Te$_7$. The half-L peak features are most pronounced at L = 6.5, where the nearby integer-L peaks are relatively weak, and the background is low. Note that the contribution from the half-lambda contamination is more than one order of magnitude weaker than the signal observed here. Such half-Bragg peaks features are not observed  for either MnBi$_2$Te$_4$ or MnBi$_6$Te$_{10}$. Figures\,\ref{Extra-1}(d) and (e)  show the rocking curves taken at half-integer Bragg peak positions for MnBi$_4$Te$_7$ (0 0 6.5) and MnBi$_6$Te$_{10}$ (0 0 34.5),  respectively, collected at both 4 K and 20 K. The strong temperature dependence (see Figures\,\ref{Extra-1}(f)) of the half-integer Bragg peaks observed in MnBi$_4$Te$_7$  excludes the possibility of half-lambda contamination and instead is indicative of a magnetic origin. The half-integer Bragg peaks observed in MnBi$_4$Te$_7$ are about 200-400 times weaker than other low-Q magnetic Bragg peaks observed along the [1 0 L] direction. These peaks are very weak and are not used for the magnetic structure refinement. However, the magnitude of the in-plane ordered moment can be estimated by comparing the intensities of (0 0 L+0.5) and (1 0 L+0.5) magnetic peaks.   This procedure yields a value of 0.15 $\mu_B$/Mn.

\subsubsection{Magnetic Structure of MnBi$_4$Te$_7$ }

\begin{table}[h]
\begin{tabular}{ccc|cccccc}
  IR  &  BV  &  Atom & \multicolumn{6}{c}{BV components}\\
        &      &             &$m_{\|a}$ & $m_{\|b}$ & $m_{\|c}$ &$im_{\|a}$ & $im_{\|b}$ & $im_{\|c}$ \\
\hline
$\Gamma_{2}$ & $\bf \psi_{1}$ &      Mn &      0 &      0 &     12 &      0 &      0 &      0  \\
$\Gamma_{6}$ & $\bf \psi_{2}$ &      Mn &      6 &      3 &      0 &      0 &      0 &      0  \\
             & $\bf \psi_{3}$ &      Mn &      0 & -5.196 &      0 &      0 &      0 &      0  \\
\end{tabular}
\caption{Basis vectors for the space group P~-3~m~1 with the magnetic wavevector ${\bf k}=( 0~ 0~0.5)$ for MnBi$_4$Te$_7$. $\Gamma_{2}$ allows ordered magnetic moments along the $c$-axis and $\Gamma_{6}$ allows ordered magnetic moments in the $ab$-plane. }
\label{Tab:Mn147bv}
\end{table}

Magnetic structure models are investigated using both the symmetry approach using the MAXMAGN program~\cite{perez2015symmetry} and the representation analysis with the program SARAh~\cite{wills2000new} to determine the symmetry-allowed magnetic structures that can result from a second-order magnetic phase transition given the crystal structure and the propagation vector. There are six possible irreducible representations (IRs) associated with the \textit{P}-3\textit{m}1  space group (\#164)and the propagation vector ${\bf {k}} = (0 0 0.5)$. The  decomposition of the magnetic representation for  the Mn $( 0 0 0.5)$ site is  $\Gamma_{Mag}=1\Gamma_{2}^{1}+1\Gamma_{6}^{2}$. The labeling of the IRs follows the scheme used by Kovalev~\cite{kovalev1993representations}. Table~\ref{Tab:Mn147bv} lists the two non-zero IRs and their associated basis vectors $\bf \psi_{i}$.  $\Gamma_{2}$ allows ordered magnetic moments along the $c$-axis and $\Gamma_{6}$ allows ordered magnetic moments in the $ab$-plane.  The weak half-integer Bragg peaks observed for MnBi$_4$Te$_7$ using HB-1A cannot be resolved at CORELLI or HB-3A due to the lower signal-to-noise ratio on these instruments. We thus refined the magnetic structure based on the representation $\Gamma_{2}$.

Figure~\ref{Fig:magstrus}(a) shows the magnetic structure model for MnBi$_4$Te$_7$ based on the representation $\Gamma_{2}$ and the refinement results on the 6~K diffraction data. This is a simple A-type antiferromagnetic order with ferromagnetic planes coupled antiferromagnetically along the $c$-axis. The refined magnitude of the ordered moments is $3.58~\pm~0.04~\mu_{B}$ per Mn ion at 6~K. Note that the quoted uncertainty only reflects the statistical error. An estimation on the systematic error using different peak integration approaches gives rise to an uncertainty on the order of 0.2~$\mu_B$ per Mn ion.

\subsubsection{Magnetic Structure of  MnBi$_6$Te$_{10}$}

\begin{table}[h]

\begin{tabular}{ccc|cccccc}
  IR  &  BV  &  Atom & \multicolumn{6}{c}{BV components}\\
      &      &             &$m_{\|a}$ & $m_{\|b}$ & $m_{\|c}$ &$im_{\|a}$ & $im_{\|b}$ & $im_{\|c}$ \\
\hline
$\Gamma_{3}$ & $\bf \psi_{1}$ &      Mn &      0 &      0 &     12 &      0 &      0 &      0  \\
$\Gamma_{5}$ & $\bf \psi_{2}$ &      Mn &      0 &     -3 &      0 &      0 &      0 &      0  \\
             & $\bf \psi_{3}$ &      Mn & -3.464 & -1.732 &      0 &      0 &      0 &      0  \\
\end{tabular}
\caption{Basis vectors for the space group R~-3~m:H with the magnetic wavevector ${\bf k}=( 0~0~1.5)$ for MnBi$_6$Te$_{10}$. $\Gamma_{3}$ allows ordered magnetic moments along the $c$-axis and $\Gamma_{5}$ allows ordered magnetic moments in the $ab$-plane.}
\label{Tab:Mn1610bv}
\end{table}

There are six possible irreducible representations (IRs) associated with the \textit{R}-3\textit{m}:H  space group (\#166) and the propagation vector ${\bf {k}} = (0~0~1.5)$. The  decomposition of the magnetic representation for  the Mn $( 0~0~0)$ site is  $\Gamma_{Mag}=1\Gamma_{3}^{1}+1\Gamma_{5}^{2}$. Table~\ref{Tab:Mn1610bv} lists the two non-zero IRs and their associated basis vectors $\bf \psi_{i}$.  $\Gamma_{3}$ allows ordered magnetic moments along the $c$-axis and $\Gamma_{5}$ allows ordered magnetic moments in the $ab$-plane.  Therefore, only $\Gamma_{3}$ agrees with the observation of no magnetic peaks along the [0 0 L] direction. From the magnetic symmetry approach, there is only one possible maximal magnetic space group, \textit{R}$_I$-3\textit{c} (\#167.108), which only allows a non-zero ordered magnetic moment along the $c$-axis, which is consistent with the magnetic structure model based on the representation $\Gamma_{3}$.

\begin{figure}[tb]
	\centering
		\includegraphics[width=0.36\textwidth]{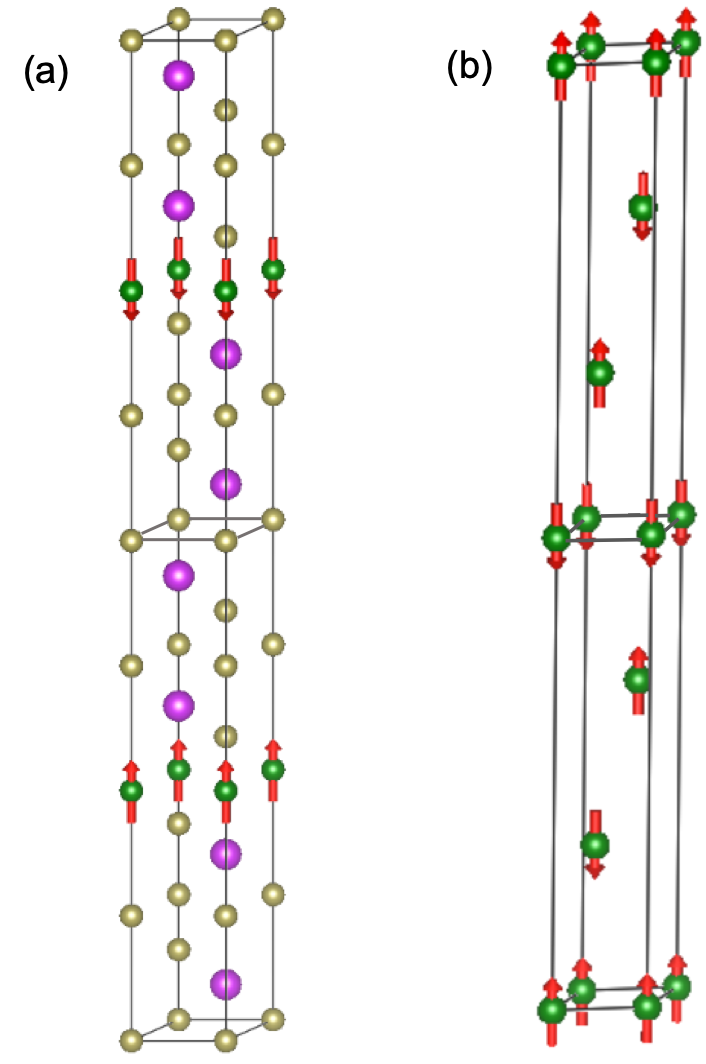}
		\caption{ \label{Fig:magstrus} (Color online) Low temperature magnetic structure determination of (a) MnBi$_4$Te$_7$ and (b) MnBi$_6$Te$_{10}$, respectively, from the neutron diffraction data and symmetry analysis. For MnBi$_6$Te$_{10}$, only the magnetic Mn atoms are shown for clarity.  In both cases, the ordered magnetic moments are along the $c$-axis in an  A-type antiferromagnetic structure with the magnetic unit cell being twice as large as the chemical unit cell along the $c$-axis.}
\end{figure}

Figure~\ref{Fig:magstrus}(b) shows the magnetic structure model for MnBi$_6$Te$_{10}$ based on the representation $\Gamma_{3}$ and the refinement results. As mentioned above, there are contributions from the tails of the nuclear Bragg peaks at the magnetic peak positions. To better separate the nuclear and the magnetic Bragg peaks, the nuclear peak integration was conducted using the 18~K dataset; while the magnetic peak integration was conducted using the difference between the 6~K and 18~K datasets, as shown in Fig.~\ref{Fig:Mn1610mesh}(c). The refined magnitude of the ordered moment is $3.55 \pm 0.04~\mu_{B}$ per Mn ion at 6~K. As for the case of MnBi$_4$Te$_7$, different peak integration approaches give rise to an uncertainty on the order of 0.2~$\mu_{B}$ per Mn ion.

\begin{table*}[!ht]

\caption{\label{str2} Summary of the properties of MnBi$_{2n}$Te$_{3n+1}$ with n=1, 2, and 3. The data for MnBi$_2$Te$_{4}$ are from Ref[\citenum{yan2019crystal, yan2019evolution}].}
\begin{tabular}{c|c|c|c}
\hline
Properties & MnBi$_2$Te$_{4}$ (n=1) & MnBi$_4$Te$_7$ (n=2)& MnBi$_6$Te$_{10}$ (n=3) \\
\hline
Space Group&\textit{R}-3\textit{m}&\textit{P}-3\textit{m}1&\textit{R}-3\textit{m}\\
a (\AA) & 4.3338(4) &4.366 & 4.374\\
c (\AA) & 40.931(6)&23.80&101.93\\
T$_N$ (K)& 24& 13& 11\\
Weiss constant (K)& 6& 11& 12\\
Effective moment ($\mu_B$/Mn) & 5.3 & 5.2& 5.0\\
Saturation moment ($\mu_B$/Mn) & 3.56 at 2K & 4.16 at 2K& 4.75 at 2K\\
Ordered moment ($\mu_B$/Mn) & 4.04(13) at 10\,K &3.6(2) at 6.5\,K& 3.6(2) at 6.5\,K\\
Interlayer coupling SJc (kOe)& 15.3& 1.5& 0.53\\
Single ion anisotropy (kOe)& 14& 17& 17\\
Field (H//c) effect & Flop at 34\,kOe, saturate at 78\,kOe & Flip at 1.6\,kOe& Flip at 1.6\,kOe\\
Field (H//ab) effect & Saturate at 103\,kOe & Saturate at 10\,kOe& Saturate at 10\,kOe\\

\hline
\end{tabular}

\end{table*}

\section{Discussion}
\subsection{Effects of stacking of the septuple and quintuple layers}

Some parameters for all three n=1, 2, and 3 members in MnBi$_{2n}$Te$_{3n+1}$ are summarized in Table III. The room temperature \textit{a} lattice parameter slightly increases with increasing n. This suggests that the stacking of the septuple and quintuple layers also modifies the in-plane competing interactions in addition to weakening the interlayer coupling. All three compounds order into an A-type antiferromagnetic structure below T$_N$ in which the magnetic moments within the triangular layers are coupled ferromagnetically and the adjacent layers are coupled antiferromagnetically. This is manifested by the anisotropic magnetic properties of MnBi$_4$Te$_7$ and MnBi$_6$Te$_{10}$ shown in Figs. \ref{chi} and \ref{MH} and further confirmed by single crystal neutron diffraction measurements. The absence of [0 0 L] magnetic reflections (Fig.\ref{Fig:Mn1610mesh}) confirms the magnetic moment is along the \textit{c}-axis for MnBi$_6$Te$_{10}$. The weak half-integer-L reflections along [0 0 L] (Fig.\ref{Extra-1}) suggest a small in-plane moment but the major conponent of the ordered moment is along the \textit{c}-axis for  MnBi$_4$Te$_7$ .

The stacking of the septuple and quintuple layers along the \textit{c}-axis has a dramatic effect on the magnetic properties in an applied magnetic field. In magnetic fields applied along the crystallographic \textit{c}-axis, the magnetic moments in MnBi$_2$Te$_{4}$ flop to an alignment perpendicular to the field above about 35\,kOe and then gradually rotate to parallel alignment above 78\,kOe. In contrast, the magnetic moments in MnBi$_4$Te$_7$ and MnBi$_6$Te$_{10}$ never become perpendicular to the applied field, but flip to be parallel to the field direction in a small critical field of about 1.6\,kOe. The latter field dependence is similar to that for MnSb$_2$Te$_{4}$ where SJc/SD is smaller than 2/z required for the spin flop transition. Here Jc is the interlayer antiferromagnetic exchange, D is the single-ion anisotropy, z is the coordination number for Mn to other Mn in layers above and below. With the magnetic septuple layers further separated by the nonmagnetic quintuple layers in MnBi$_4$Te$_7$ and MnBi$_6$Te$_{10}$, the interlayer coupling SJc is reduced while the single ion anisotropy SD is less affected.

As for the case of MnSb$_2$Te$_{4}$\cite{yan2019evolution}, the interlayer antiferromagnetic exchange (Jc) and single-ion anisotropy (D) can be estimated from the critical magnetic field H$_c$ for the spin flip transition and the critical field H$_c^*$ applied perpendicular to the \textit{c}-axis to saturate the magnetic moment: SJc=g$\mu_B$H$_{c}$/z, SD=(1/2)g$\mu_B$H$_c^*$-zSJc, where g=2. It is worth mentioning that z=2 for MnBi$_4$Te$_7$ and z=6 for MnBi$_6$Te$_{10}$. As expected, SJc=1.5\,kOe of MnBi$_4$Te$_7$ is about three times of SJc=0.53\,kOe for MnBi$_6$Te$_{10}$; SD is comparable to each other for these two compounds. It should be noted that SD is also comparable to those in MnSb$_2$Te$_{4}$ and MnBi$_2$Te$_{4}$. These parameters are listed in Table III. The ratio of SJc/SD is 0.2 and 0.08 for MnBi$_4$Te$_7$ and MnBi$_6$Te$_{10}$, respectively. Both are smaller than 1/3 required for the occurrence of a spin flop transition. The spin gap estimated as $\bigtriangleup$=2SD$\sqrt{zSJc/SD+1}$ is around 37\,kOe for both compounds.

The above estimation suggests that the separation of the magnetic septuple layers by the nonmagnetic  quintuple layers in MnBi$_{2n}$Te$_{3n+1}$ with increasing n significantly reduces the interlayer coupling SJc. However, the experimentally observed A-type antiferromagnetic order with T$_N>$10\,K indicates an important role of SJc in determining the magnetic structure. In MnBi$_6$Te$_{10}$ with n=3, SJc, although only about 8\% of SD, seems to be enough to stabilize the A-type magnetic order. However, with further increasing the number of the nonmagnetic quintuple layers in between the magnetic septuple layers, SD is expected to stabilize a long range ferromagnetic order\cite{binek2003ising}. It will be interesting to synthesize and check the magnetic properties of MnBi$_{2n}$Te$_{3n+1}$ with n$\geq $4. The persistence of magnetism in MnBi$_{2n}$Te$_{3n+1}$  series will be further discussed later.

Despite the dramatic effect of the stacking of the septuple and quintuple layers on the magnetic properties, all three compounds show similar temperature dependence of transport properties. Cooling across T$_N$, the in-plane electrical resistivity shows a cusp-like feature; the in-plane thermal conductivity shows a dip-like feature typical of critical scattering; and thermopower shows no detectable anomaly. The dip-like feature induced by the critical scattering is less pronounced in MnBi$_4$Te$_7$ and MnBi$_6$Te$_{10}$ as shown in Fig.\,\ref{kappa}(b). This might result from the dilute effect by the nonmagnetic quintuple layers. This dilution effect can also contribute to the weaker lambda-type anomalies in specific heat (see Fig.\,\ref{Cp-1}) with increasing n.

The stacking of the septuple and the quintuple layers along the crystallographic \textit{c}-axis can also affect the topological properties in MnBi$_{2n}$Te$_{3n+1}$ compounds. In MnBi$_2$Te$_{4}$, the termination surface is always the septuple layers and recent photoemission studies reported gapless Dirac surface states  \cite{hao2019gapless,chen2019topological,li2019dirac,swatek2019gapless}. In MnBi$_4$Te$_7$ and MnBi$_6$Te$_{10}$, the termination surface can be the septuple layers or the quintuple layers. Recently photoemission studies on MnBi$_4$Te$_7$ reported gapped (001) surface states when the termination surface is the septuple layer; when the termination layer is the quintuple layer, the upper Dirac cone goes deeper down in energy and merges with the bulk valence band\cite{hu2020van,wu2019natural}. While the origin and observation of the gapless Dirac surface states in MnBi$_2$Te$_{4}$ are still under debate, it has been proposed that the surface septuple layer might show a different magnetic order than the bulk\cite{hao2019gapless,swatek2019gapless}. In MnBi$_4$Te$_7$, the quintuple layer beneath the termination septuple layer can help isolate the termination septuple layer from the bulk. In the case that the termination layer is the nonmagnetic quintuple layer, it can behave as a capping layer and reduce the relaxation or reconstruction of the magnetic septuple layer. From this point of view, it would be interesting to check the band structure of MnBi$_6$Te$_{10}$ with the two quintuple layers on top of the septuple layer.

\subsection{Persistence of magnetism}

It is interesting to notice the {\it persistence} of the
magnetism, both in terms of ordered Mn moments of order 4 $\mu_B$, as well as significant Neel temperatures above 10 K, even as the \textit{c}-axis distance between
adjacent Mn layers reaches dozens of Angstroms as in MnBi$_6$Te$_{10}$. Given the general exponential fall-off (at large distances) of atomic-level wavefunctions such as those relevant for Mn (local) magnetism, should not the interlayer exchange coupling rapidly fall to zero as the number of intervening quintuple layers increases, thereby driving the Neel point to zero, along with the ordered moment? It is {\it a priori} rather surprising for a material, such as MnBi$_6$Te$_{10}$, in which only one of 17 atoms is a characteristically magnetic atom, to exhibit a three-dimensional ordered spin structure with a Neel temperature over 10 K as found here.

In order to understand the persistence of magnetism in MnBi$_{2n}$Te$_{3n+1}$ with n$\geq$2, we have performed first principles calculations. These new results, along with the previous Monte Carlo theoretical work of Yasuda et al \cite{yasuda2005neel}, provide a partial answer to this question. For each of the three compounds MnBi$_2$Te$_4$, MnBi$_4$Te$_7$ and MnBi$_6$Te$_{10}$, we have used the augmented plane-wave density functional theory code WIEN2K \cite{blaha2001wien2k}, employing the generalized gradient approximation of Perdew, Burke and Ernzerhof \cite{perdew1996generalized}. We have used a value of 8.0 for RK$_{max}$, in general the product of the smallest muffin-tin radius and the largest plane-wave expansion wavevector. For all calculations and all atoms a muffin-tin radius of 2.5 Bohr was used.  Experimental lattice parameters were used, with internal coordinates optimized in a ferromagnetic configuration, as substantial recent work \cite{pokharel2018negative, chen2019suppression} in this area has found strong magnetoelastic coupling in these materials. Spin-orbit coupling was not included.

For MnBi$_2$Te$_4$, MnBi$_4$Te$_7$, we have also calculated the interlayer magnetic coupling, which is defined here as simply the energy difference, per Mn, between the ferromagnetic state and a state with neighboring Mn planes anti-aligned. For MnBi$_6$Te$_{10}$, this last calculation did not yield reliable results (possibly related to the highly anisotropic crystal structure) and we will therefore discuss the interlayer coupling here in terms of the experimental results. In any case the interlayer coupling for MnBi$_6$Te$_{10}$ is likely sufficiently small as to be near the accuracy of the calculation, given the Mn-Mn interlayer distance here of over 30 \AA, so it is sensible to discuss this based on the experimental results.

For all materials here our calculations find the Mn to be in a high-spin ($S = 5/2$) state, as for all three materials, the ordered ferromagnetic moment (per Mn) is 5.0$\mu_B$.  The ordered moment in the experimentally observed A-type
antiferromagnetic ground state is smaller, but this value is only within the muffin-tin spheres (there is a similar muffin-tin-only value for the ferromagnetic state). For these two materials this ground state has insulating character, with respective band gaps of 0.63 and 0.45 eV. These band gaps may be understated relative to actual values given the usual GGA underestimation of band gaps. We believe that the smaller experimentally observed  ordered moment may be due to the presence of some disorder in the sample, given the partial occupancies that several structural refinements in the literature find in this family of materials.

The calculated interlayer coupling, as discussed above, is 3 meV/Mn for MnBi$_2$Te$_4$ and 0.5 meV for MnBi$_4$Te$_7$.  This 6:1 ratio is in reasonable accord, considering the disorder,  with the ratio of nearly 10:1 derived from magnetic data (See Table III). Following the results listed in Table III, we will simply assume that the interlayer coupling for the MnBi$_6$Te$_{10}$ compound is 1/3 of the calculated value for the MnBi$_{4}$Te$_{7}$ compound, or 0.17 meV/Mn.  This is less than 2 K and is substantially smaller than the 11 K ordering temperature. In the simplest isotropic approximation (clearly inapplicable here), ordering temperatures are estimated as a third of these energy differences, or less than 1 K. How can all these materials then exhibit ordering temperatures much larger than these values?

We provide a possible explanation for this behavior in terms of previous Monte Carlo work on quasi-two dimensional Heisenberg antiferromagnets \cite{yasuda2005neel}, which focused specifically on the effects on the Neel point T$_{N}$ of the ratio of exchange constants J$^{'}/$J, where in this case J$^{'}$ would be the interlayer exchange constant and J the planar exchange constant.  Note that that theoretical work assumed a simple cubic lattice, while here we have highly anisotropic hexagonal and rhombohedral crystal structures, so that we cannot claim exact applicability of these results here.  Nevertheless, the basic physics of anisotropic magnetism is likely to be similar.  

The most relevant finding from that work is that for small values J$^{\prime}/J \ll 1$, the Neel point follows the approximate relationship $T_{N} \sim -1/\ln(J^{\prime}/J)$.  Now, this is in fact a very weak relationship for small $J^{\prime}$; in particular if the ratio J$^{\prime}/J$ is allowed to vary from 0.1 to 0.01 to 0.001, the corresponding ratios of Neel points would be approximately 1:1/2:1/3. Thus a hundred-fold reduction in interlayer exchange coupling only reduces the ordering temperature by an approximate factor of 3.  This is again broadly consistent with our experimental finding that a thirty-fold reduction in interlayer coupling (from MnBi$_2$Te$_4$ to MnBi$_6$Te$_{10}$) only reduces the ordering temperature by roughly a factor of 2.

Finally, we briefly discuss the theoretical finding that the ordered moment is 5.0 $\mu_B$ for all three materials. This is the expected value for divalent Mn in an insulating material, and mainly reflects the {\it local} physics of moment formation, as opposed to the ordering temperature, which derives from {\it interatomic} exchange interactions. These interactions, for the interlayer coupling, must necessarily involve numerous intervening atoms.  Yet in all these structures, there are 6 Te nearest-neighbor atoms to Mn, just as in MnTe itself, which is also an $S= 5/2$ material. Hence a picture emerges in which the main function of the intervening Bi$_2$Te$_3$ layers is to reduce the interlayer exchange coupling and thereby the Neel temperature, though we have argued that the fall-off of Neel temperature with this coupling is rather weak. Since compounds with even more intervening Bi$_2$Te$_3$ layers are possible\cite{amiraslanov2018}, the argument presented suggests that these too may have significant ordering temperatures.

\section{Summary}
In summary, we study the A-type antiferromagnetic order in MnBi$_4$Te$_7$ and MnBi$_6$Te$_{10}$ single crystals by measuring the magnetic, transport, and thermodynamic properties as well as single crystal neutron diffraction. Both MnBi$_4$Te$_7$ and MnBi$_6$Te$_{10}$ order into an A-type antiferromagnetic structure with ferromagnetic layers coupled antiferromagnetically. A small in-plane ordered moment in MnBi$_4$Te$_7$ is identified through the detection of weak half-integer magnetic Bragg reflections along the [0 0 L] direction. The comparison with MnBi$_2$Te$_4$ suggests that the stacking of the magnetic septuple layers and increased spacing between them have dramatic effect on the magnetic properties. The nonmagnetic quintuple layers between the magnetic septuple layers significantly reduce the magnitude of interlayer coupling. However, single ion anisotropy shows little change with increasing n in MnBi$_{2n}$Te$_{3n+1}$ compounds. This suggests that the septuple layers are similar in all these compounds. This also indicates that a long range ferromagnetic order is expected when further increasing the spacing between the septuple layers.

MnBi$_4$Te$_7$ and MnBi$_6$Te$_{10}$ single crystals grown out of Bi-Te flux as reported in this work tend to be Mn-deficient and Bi-rich. Mn deficiency and Mn/Bi site mixing can affect the magnetic and transport properties and should be considered in future work optimizing the growth parameters and physical properties. With more complex stacking of the magnetic septuple and the nonmagnetic quintuple layers in counmpounds with n$>$1, the formation of stacking faults becomes more likely. The local stacking faults might not dramatically modify the long range magnetic order of the crystal. However, they can become critical for devices made out of thin flakes.

\section{Acknowledgment}

The authors would thank H. Cao, A. F. May, R. J. McQueeney, S. Okamoto, Q. Zhang for helpful discussions. Work at ORNL was supported by the U.S. Department of Energy, Office of Science, Basic Energy Sciences, Materials Sciences and Engineering Division. A portion of this research used resources at the Spallation Neutron Source and High Flux Isotope Reactor, which are DOE Office of Science User Facilities operated by the Oak Ridge National Laboratory.

\section{references}

%

\end{document}